\documentclass[sigconf]{acmart}

\renewcommand\footnotetextcopyrightpermission[1]{} 
\setcopyright{none}

\acmDOI{}

\acmISBN{}

\acmConference[Preprint 2019]{Conference preprint}{2019}{Preprint, Preprint}
\acmYear{2019}
\copyrightyear{2019}

\acmArticle{4}
\acmPrice{15.00}

\usepackage{amsmath} 
\usepackage{algorithm,amsmath} 
\usepackage[noend]{algorithmic}
\usepackage{booktabs}
\usepackage{enumitem}
\usepackage{subcaption}
\usepackage{tabularx}
\usepackage{textcomp} 
\usepackage{CJKutf8} 

\usepackage{float}
\floatstyle{plaintop}
\restylefloat{table}

\usepackage{tikz}
\newcommand{\rightanglearrow}[1][]{%
  \begin{tikzpicture}[#1]%
    \draw (0,0.7ex) -- (0,0) -- (0.75em,0);
    \draw (0.55em,0.2em) -- (0.75em,0) -- (0.55em,-0.2em);
  \end{tikzpicture}%
}

\usepackage{draftwatermark}
\SetWatermarkText{Draft}
\SetWatermarkScale{1}
\SetWatermarkLightness{0.9}

\definecolor{officeblue}{RGB}{38, 68, 120}
\definecolor{officeviolet}{RGB}{112, 48, 160}
\definecolor{officeorange}{RGB}{197, 90, 17}

\newcommand\boldofficeblue[1]{\textcolor{officeblue}{\textbf{#1}}}

\newcommand\boldunderofficeblue[1]{\textcolor{officeblue}{\underline{\smash{\textbf{#1}}}}}
\newcommand\boldunderofficeviolet[1]{\textcolor{officeviolet}{\underline{\smash{\textbf{#1}}}}}

\newcommand\boldunder[1]{\underline{\smash{\textbf{#1}}}}

\newcommand{\cmmnt}[1]{\ignorespaces}

\newcommand{\rowfonttype}{}
\newcommand{\rowfont}[1]{
   \gdef\rowfonttype{#1}#1%
}
\newcolumntype{L}{>{\rowfonttype}l}

\newcolumntype{Y}{>{\centering\arraybackslash}X}

\begin{document}


\title{IP Geolocation through Reverse DNS}

\author{Ovidiu Dan}
\orcid{0000-0001-5712-3544}
\affiliation{%
  \institution{Lehigh University}
  \city{Bethlehem}
  \state{PA}
  \country{USA}
  \postcode{18015}
}
\additionalaffiliation{%
  \institution{Microsoft Bing}
  \city{Redmond}
  \state{WA}
  \country{USA}
  \postcode{98052}
}
\email{ovd209@cse.lehigh.edu}

\author{Vaibhav Parikh}
\affiliation{%
  \institution{Microsoft Bing}
  \city{Redmond}
  \state{WA}
  \country{USA}
  \postcode{98052}
}
\email{vparikh@microsoft.com}

\author{Brian D. Davison}
\orcid{0000-0002-9326-3648}
\affiliation{%
  \institution{Lehigh University}
  \city{Bethlehem}
  \state{PA}
  \country{USA}
  \postcode{18015}
}
\email{davison@cse.lehigh.edu}

\begin{abstract}
IP Geolocation databases are widely used in online services to map end user IP addresses to their geographical locations. However, they use proprietary geolocation methods and in some cases they have poor accuracy. We propose a systematic approach to use publicly accessible reverse DNS hostnames for geolocating IP addresses. Our method is designed to be combined with other geolocation data sources. We cast the task as a machine learning problem where for a given hostname, we generate and rank a list of potential location candidates. We evaluate our approach against three state of the art academic baselines and two state of the art commercial IP geolocation databases. We show that our work significantly outperforms the academic baselines, and is complementary and competitive with commercial databases. To aid reproducibility, we open source our entire approach.
\end{abstract}



\fancyfoot{}
\maketitle
\thispagestyle{empty}



\section{Introduction}
\label{Sec:Introduction}




IP Geolocation databases map IP addresses to their corresponding geographical locations. They are often used to find the approximate location of an IP address at the city level. Records in these databases typically contain IP ranges along with their physical location. These databases are vital to a variety of online services when the exact location of a user is not available. Table \ref{table:Example of rows from IP Geolocation DB} lists a few examples of such records. For instance, the second example in the table maps a /24 subnet (256 IPs) to Hengyang, a city in China. While some users opt-in to share their exact coordinates to online services through devices with global positioning sensors or Wi-Fi based geolocation, others decline or use devices without such features. IP geolocation is therefore a valuable source of information on user location. 



\begin{figure*}[!t]
	\centering

	\begin{subfigure}{.45\textwidth}
  		\centering
		\frame{\includegraphics[height=5cm]{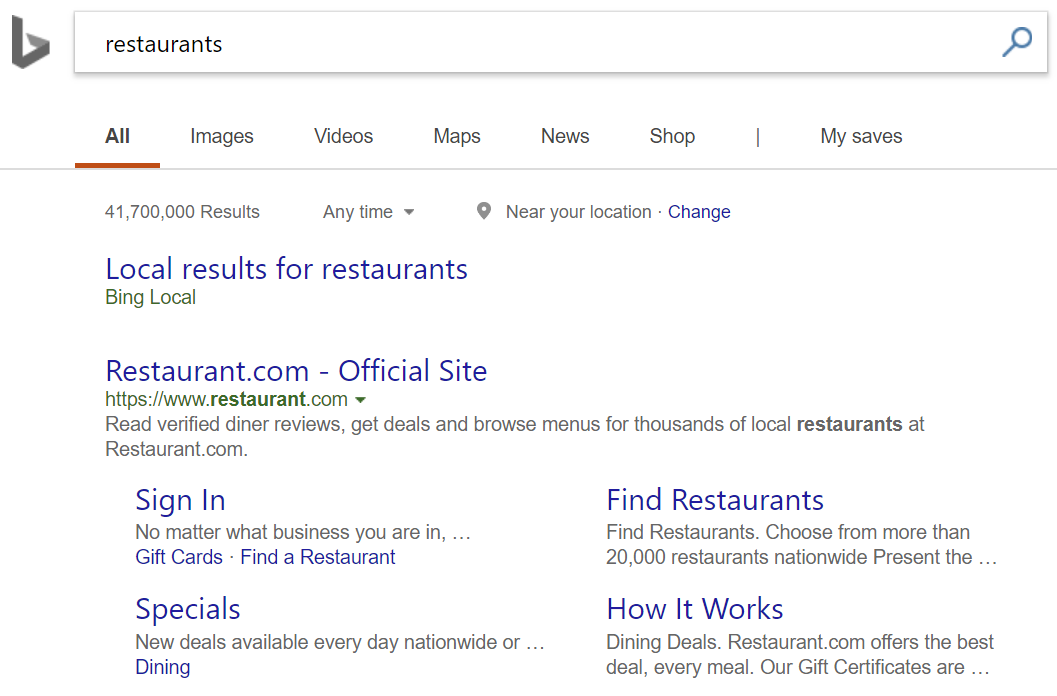}}
		\label{fig:sub1}
	\end{subfigure}%
	\begin{subfigure}{.55\textwidth}
		\centering
		\frame{\includegraphics[height=5cm]{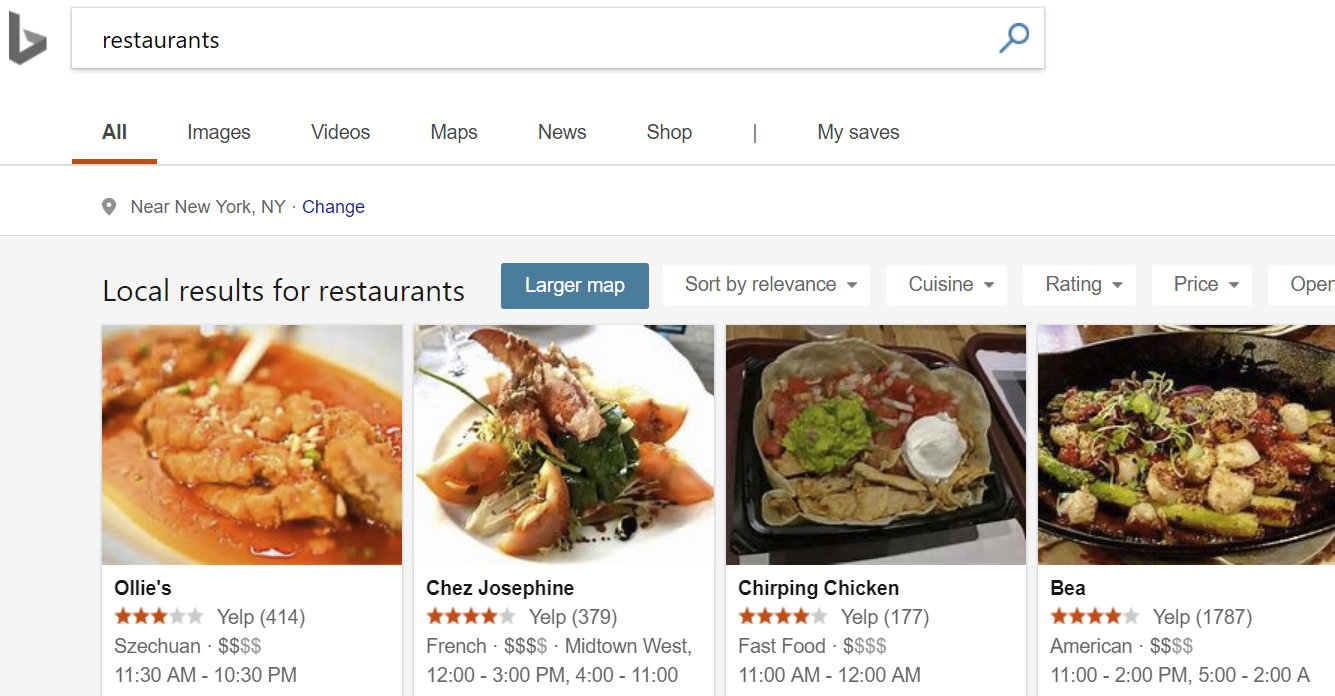}}
		\label{fig:sub2}
	\end{subfigure}
    
    \caption{Effect of missing location information for the query "restaurants". The left image displays the results when the location is unknown. The right image displays a personalized experience for a user located in New York City.}
	\label{fig:Effect of missing location on search results}
\end{figure*}

A practical application of IP geolocation is \textbf{personalized local search results} in the context of search engines. Figure \ref{fig:Effect of missing location on search results} demonstrates the striking difference in results for the query "restaurants" when the user location is unknown, compared to when it is known. The generic nationwide results require the user to requery for more specific restaurants in their area, while the personalized results directly list restaurants tailored to a specific location. Previous work has shown that personalizing results to a user's location leads to increased user satisfaction and conversely that missing location information leads to user dissatisfaction \cite{dan2016improving, bennett2011inferring, kliman2015location}. IP geolocation databases are also used in many other applications, including: \textbf{content personalization and online advertising} to serve content local to the user \cite{bennett2011inferring, hannak2013measuring, kolmel2002location}, \textbf{content delivery networks} to direct users to the closest datacenter \cite{huang2011public}, \textbf{law enforcement} to fight cybercrime \cite{Shue2013}, \textbf{geographic content licensing} to restrict content delivery to licensed geographic regions \cite{macvittie2010geolocation}, and \textbf{e-commerce} to display variable pricing based on local taxes and shipping \cite{svantesson2007commerce}.

\begin{figure}[htb]
\centering
\includegraphics[width=0.8\linewidth]{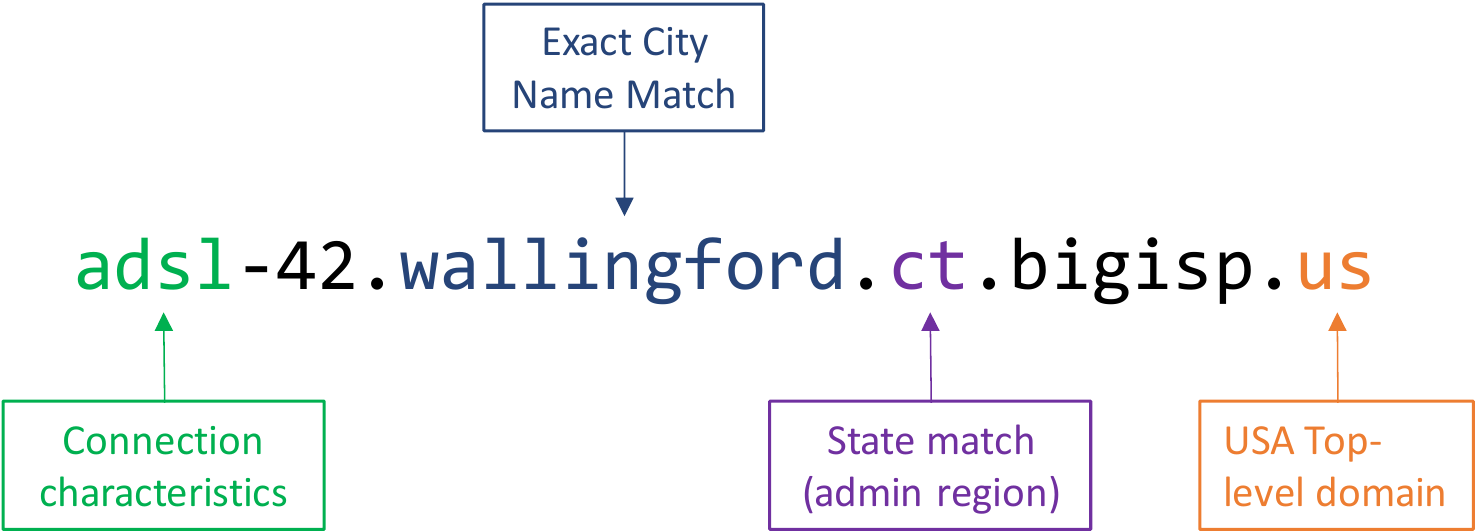}
\caption{Example of information that can be extracted from reverse DNS hostnames, including location information such as city name, state, country, as well as physical connection characteristics.}
\label{fig:ReverseDNSExample1}
\end{figure}

\begin{table}
\footnotesize
\caption{Example of entries from an IP Geolocation database}
\label{table:Example of rows from IP Geolocation DB}
\begin{tabularx}{\columnwidth}{ L L L L L }
    \toprule
    \rowfont{\footnotesize}%
    \textbf{StartIP} & \textbf{EndIP} & \textbf{Country} & \textbf{Region} & \textbf{City} \\
    \midrule
    1.0.16.0   & 1.0.16.255   & JP      & Tokyo        & Tokyo     \\
    124.228.150.0   & 124.228.150.255   & CN      & Hunan    & Hengyang \\
    131.107.147.0 & 131.107.147.255 & US      & Washington & Redmond	\\
	\bottomrule
  \end{tabularx}
\end{table}

Commercial IP geolocation databases such as MaxMind \cite{maxmind2018}, Neustar IP Intelligence \cite{neustar2018}, and IP2Location \cite{ip2location2018} are considered state of the art. 
They combine multiple IP location  sources, including WHOIS lookups, network latency, network topology, reverse DNS, as well as direct contracts with Internet Service Providers \cite{muir2009internet}. However, the exact methods they use are proprietary. Some of these approaches have been studied in academia to some extent, as described in Section \ref{sec:Related Work}. Related work has shown that while they have high coverage, commercial databases are sometimes inaccurate or are missing location information for some IP ranges \cite{shavitt2011geolocation,poese2011ip,gharaibeh2017look}. 


\textbf{Our work focuses on extracting location information from reverse DNS hostnames assigned to IP addresses}, which has many potential advantages including high coverage and accuracy. 
These hostnames can be periodically collected in a short amount of time by performing a reverse DNS lookup for every address in the IP space. Figure \ref{fig:ReverseDNSExample1} exemplifies the information that can be parsed from reverse DNS hostnames. Here we can derive both the location and connection characteristics for the hostname of an IP address. A person reading the name of the hostname can reasonably determine that it references \textit{Wallingford}, a town in Connecticut, USA. 

\textbf{Reverse DNS is the opposite of Forward DNS}. Forward DNS starts from a domain or subdomain such as \texttt{www.bing.com} and resolves to zero, one, or more IP addresses \cite{rfc1034}. Note that multiple subdomains can map to the same IP. Conversely, reverse DNS lookups start from an IP address and typically return zero or one hostnames \cite{rfc2317}. The reverse DNS hostname does not need to be the same as the Forward DNS hostname. While Forward DNS lookups are used by Internet users to get to websites, Reverse DNS hostnames are typically used to name and describe the underlying physical infrastructure that makes up the Internet.

\begin{figure}[h]
	\centering
    {
		\includegraphics[width=0.4\textwidth]{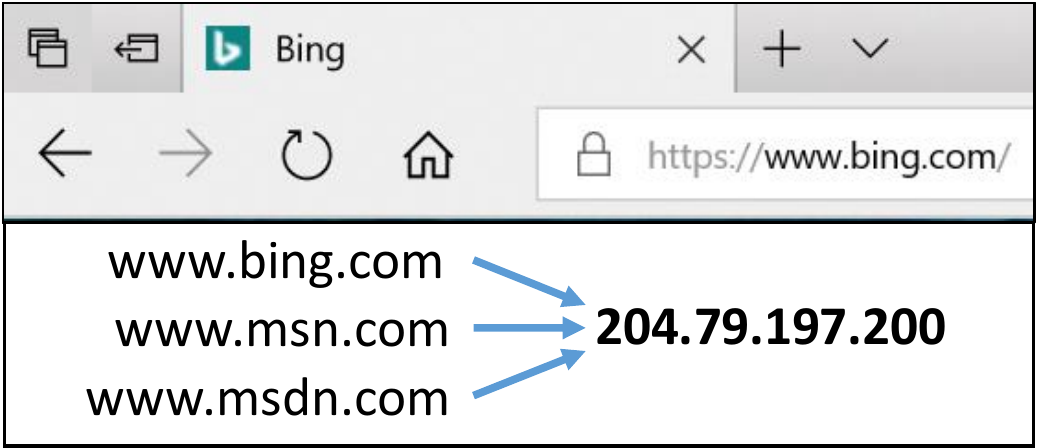}%
	}\par\medskip
    {
		\includegraphics[width=0.4\textwidth]{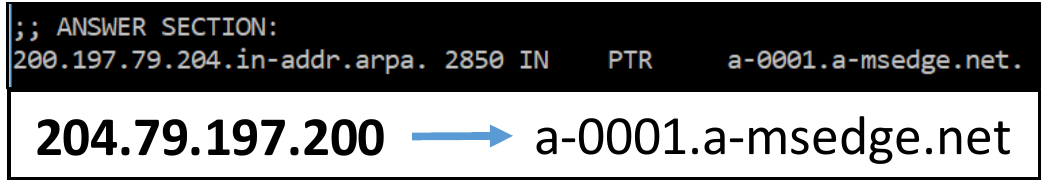}%
	}\par\medskip
    \caption{Example of the difference between Forward DNS (top) and Reverse DNS (bottom)}
	\label{fig:ForwardReverseDNS}
\end{figure}

Figure \ref{fig:ForwardReverseDNS} contains examples of both forward and reverse DNS resolution. 

\textbf{Given a reverse DNS hostname, our task is to determine its location at the city level}. 
This task poses multiple challenges. First, the naming schemes of Internet Service Providers are often ad-hoc and do not always contain the full names or common abbreviations of cities. For example, the \texttt{drr01.cral.id.frontiernet.net} hostname is located in \textit{Coeur D'Alene, Idaho}. Determining that the \texttt{cral} substring maps to this location is difficult even for a human. Second, many cities around the world have ambiguous names. Take for instance \textit{Vancouver, Canada} and \textit{Vancouver,  USA}. A hostname which only contains the substring \textit{vancouver} is not specific enough to determine a single location correctly. Even unambiguous city names can become ambiguous when abbreviations are used instead of their full names. Does \texttt{nwmd} refer to \textit{\textbf{N}e\textbf{w} Rich\textbf{m}on\textbf{d}, WI} or to \textit{\textbf{N}e\textbf{w} \textbf{M}arylan\textbf{d}, NB}, or to neither of them? Third, sometimes hostnames contain multiple or conflicting locations. For example, it is difficult to determine if \texttt{sur01.tacoma.wa.seattle.comcast.net} is located in \textit{Seattle, WA}, \textit{Tacoma, WA}, or maybe even \textit{\textbf{Su}mne\textbf{r}, WA}.




We propose a systematic approach for using reverse DNS hostnames to geolocate IP addresses. Our contributions are:

\begin{enumerate}[topsep=2pt,leftmargin=10pt,labelsep=*]


  \item \textbf{In our preliminary investigation} we determine reverse DNS coverage in the entire IPv4 address space. We also find an upper bound of exact city and airport code matches.

  \item \textbf{We present a machine learning approach for extracting locations from hostnames}. We cast the task as a machine learning problem where for a given hostname, we split the hostname into its constituent terms, we generate a list of potential location candidates, and then we classify each hostname and candidate pair using a binary classifier to determine which candidates are plausible. Finally, we rank the remaining candidates by confidence, and we break the ties by location popularity.

  \item \textbf{We evaluate our approach against state-of-the-art baselines}. Using a large ground truth set, we evaluate our approach against three academic baselines and two commercial IP geolocation databases. We show that our method significantly outperforms academic baselines. We also show that the academic baselines contain incorrect rules which impact their performance. Finally, we demonstrate that our approach is both competitive and complementary to commercial geolocation baselines, which shows that our method can help improve their accuracy.
 
    \item \textbf{We release our approach as open source}. To help the academic community reproduce our results, we release our reverse DNS geolocation software as open source.

\end{enumerate}


\section{Related Work}
\label{sec:Related Work}

We divide IP geolocation research in two broad categories, based on the methods they use: \textbf{network delay and topology} approaches use ping, traceroute, and BGP network structure information; \textbf{web mining} approaches use diverse information mined from the web, including WHOIS databases, social graphs, and reverse DNS.

The majority of IP geolocation research relies on active \textbf{network delay measurements} to locate addresses. These approaches issue pings or traceroutes from multiple geographically distributed landmark servers to a target IP, then triangulate the location \cite{padmanabhan2001investigation,jayant2004toward, gueye2006constraint, katz2006towards, youn2009statistical, laki2010model, eriksson2010learning, li2012ip, ciavarrini2018geolocation}. Multiple systems such as Octant \cite{wong2007octant}, Alidade \cite{chandrasekaran2015alidade}, or HLOC \cite{scheitle2017hloc} combine delay measurement methods with other data sources such as reverse DNS and WHOIS information.

Network delay and topology methods have significant limitations. First, they have scalability issues as they require access to landmark servers distributed around the world and each target IP needs separate measurements. Second, not all networks allow ping and traceroute. Third, routes on the Internet do not necessarily map to geographic distances. These problems often lead to lackluster results, with error distance in the order of hundreds of kilometers \cite{padmanabhan2001investigation,gueye2006constraint}. Some of the earlier research is also plagued by extremely small ground truth datasets, often focusing on a handful of IP addresses in a few US universities \cite{padmanabhan2001investigation, katz2006towards, gueye2006constraint, youn2009statistical}.

Our work addresses several of these limitations. First, using reverse DNS hostnames for geolocation does not require any network delay measurements. Reverse DNS hostnames can be obtained much faster than performing active delay measurements, by querying DNS servers. Second, our ground truth dataset is several orders of magnitude larger than the ones used in previous work and it spans the entire planet. Third, our approach of extracting locations from hostnames can be performed offline.
Fourth, our results have lower median error distance than most previous work.


\textbf{Web mining approaches} use diverse information mined from the web. Guo et al. \cite{guo2009mining} extract locations mentioned in web pages and assign the locations to the IPs which host the content. Although they report city level agreement for 87\% of the IPs, they use an IP geolocation database with unknown accuracy as ground truth. Endo and Sadok \cite{endo2010whois} propose using \textit{whois} information. Unfortunately, the evaluation section lacks a comparison against ground truth. Wang et al.~\cite{wang2011towards} combine a network delay approach with extracting the location of web servers from the web pages that they host. They obtain a median error of 0.69 kilometers for the best data set, which contains only 88 target IP addresses. Backstrom et al.~\cite{backstrom2010find} propose deriving the location of target users based on the locations of friends. However, this approach requires access to users' social graphs. It achieves a median error distance of 590 km on 2,830 IPs.

\textbf{In this work we propose extracting IP locations from their reverse DNS hostnames}. 
\textit{GeoTrack} \cite{padmanabhan2001investigation}, proposed by Padmanabhan and Subramanian, is one of the earliest reverse DNS geolocation approaches. It uses manual rules to determine locations of hostnames in the United States using city names, airport codes, and country codes. It then combines this approach with traceroutes to estimate the location of a target IP. Our machine learning approach does not require manual rules and it achieves a median error of tens of kilometers using a test set of millions of IP addresses.






\textit{Undns} is the most well-known and widely used reverse DNS geolocation approach \cite{undns2002}. Similarly to \textit{GeoTrack}, it consists of manual rules which are expressed as regular expressions at the domain level. For example, the rule \texttt{([A-Z]{3,4})[0-9]?.verizon-gni.net} matches the hostname \texttt{PHIL.verizon-gni.net}. A domain specific location dictionary is then used to match the extracted slot \texttt{PHIL} to \textit{Philadelphia, PA}. The obvious disadvantage of this approach is that each domain requires manually generated and potentially error prone rules. In comparison, our approach is more scalable since it does not require human generated rules. It also handles unique situations better, since it considers the terms of each hostname individually, without requiring domain-specific training. In Section \ref{sec:Academic baselines} we show that our approach significantly outperforms \textit{undns}. Multiple geolocation and network topology papers use \textit{undns} as-is to draw conclusions or perform experiments \cite{freedman2005geographic,2018nurgeography}. Unfortunately, we demonstrate that \textit{undns} results suffer due to catch-all rules.

\textit{DRoP}, another state of the art reverse DNS based approach, aims to geolocate hostnames using automatically generated rules generated by finding patterns across all the hostname terms of a domain \cite{drop2014}. For example, it may find that for the domain \texttt{cogentco.com}, the second term from the right often contains airport codes. These rules are then validated using network delay information. \textit{DRoP} places 99\% of IPs in 5 test domains within 40 km of their actual location. However, it uses network delay measurements, its method of splitting hostnames is rudimentary, and as we show in Section \ref{sec:Academic baselines}, it performs poorly on worldwide ISP domains.

Finally \textit{DDec} \cite{ddec2018} combines \textit{undns} and \textit{DRoP} rules by giving precedence to \textit{undns} and using \textit{DRoP} as fallback.

\section{Datasets}
\label{sec:Datasets}

This section contains descriptions of the datasets we use throughout this paper for experiments, training and testing.

\textbf{Our ground truth set} contains 67 million IP addresses with known geographic location. To the best of our knowledge, it is the largest and most diverse set used in geolocation literature. We compiled the ground truth set in March 2018 by randomly sampling the query logs of a large-scale commercial search engine. We describe the characteristics of this dataset in more detail in Section \ref{sec:Ground Truth}.

\textbf{GeoNames} is a free database with geographical information \cite{geonames2018}. The March 2018 snapshot we used contains information on 11.5 million geographic features from all countries in the world. From Geonames we used multiple subsets available separately for download. \textit{Cities 1000} consists of information on all cities in the world with a population of at least 1,000, including coordinates, original names, ASCII names, alternate names, and codes of administrative divisions. \textit{Alternate Names} contains more alternate names for some cities such as abbreviations, colloquial names, and historic names. More importantly, it also contains \textit{airport codes} issued by \textit{IATA}, \textit{ICAO}, and \textit{FAAC}, which are travel organizations. \textit{Admin 1 Codes} is comprised of the codes and names of first-level administrative regions. \textit{Country Info} contains general information about countries, including their Internet top-level domain (TLD). 

\textbf{CLLI} is an abbreviation for Common Language Location Identifier. These codes are used by the North American telecommunications industry to designate names of locations and functions of telecommunications equipment. While historically only used by the Bell Telephone companies, they were more recently adopted by other companies as well. Multiple codes can map to the same location. For example, all the following codes map to \textit{Chicago, Illinois}: \texttt{chcgil}, \texttt{chchil}, \texttt{chciil}, \texttt{chcjil}, and \texttt{chclil}. Note that the codes cannot necessarily be derived from the name of the city. This database is available from multiple sources. We acquired a May 2017 snapshot from TelcoData \cite{telcodata2018} for a token amount.

\textbf{UN/LOCODE}, which stands for United Nations Code for Trade and Transport Locations, is a worldwide geographic coding scheme developed and maintained by the UN. It assigns codes to locations used in trade and transport, such as rail yards, sea ports, and airports. The code assigned to Paris, France is \texttt{FRPAR} and the functions listed for this location are: \texttt{port}, \texttt{rail}, \texttt{road}, and \texttt{postal}. This dataset is updated twice a year and it is available for free on the United Nations Economic Commission for Europe website \cite{unlocode2018}. We used the December 2017 release, which was the latest available version.

\textbf{Public Suffix List}, maintained by the Mozilla Foundation, is a free list of domain suffixes under which Internet users can directly register names \cite{publicsuffix2018}. Examples include \texttt{cloudapp.net} and \texttt{gov.uk}.

\textbf{Rapid7 Reverse DNS} consists of reverse DNS hostnames of the entire IPv4 address space. The dataset is available for free and it is updated weekly. The archive contains snapshots going back to 2013 \cite{rapid7rdns2018}. We discuss this dataset in detail in the next section.

\section{Reverse DNS}
\label{sec:Reverse DNS}

Forward DNS lookups convert hostnames such as \texttt{www.bing.com} into IP addresses, while reverse DNS lookups work in the other direction; they start from an IP address and find a 
hostname. \textbf{Since the forward and reverse DNS lookups are set in different DNS records, they do not need to have the same hostname.} Reverse hostnames are more likely to be used to name the underlying networking infrastructure, while forward hostnames are used to name websites or other online services \cite{rfc1034}. 




IPv6 addresses also have reverse DNS hostnames. The only difference is that the records are under the \texttt{ip6.arpa} domain. \textbf{While we evaluate our approach on IPv4, all methods described in this paper can be equally applied to IPv6 addresses as well.}

To determine the viability of using reverse DNS hostnames for geolocation, we studied the \textit{Rapid7 reverse DNS dataset} \cite{rapid7rdns2018}, which covers the entire IPv4 address space. Rapid7 compiles it by performing IPv4 PTR lookups over the entire address space as described above, except for ranges that are blacklisted or private. The archive contains snapshots going back to 2013 \cite{rapid7rdns2013}. This preliminary investigation  is based on a snapshot taken in January 2017, while in the rest of the paper we use a more recent dataset from March 2018.

The IPv4 address space consists of all 32-bit numbers. This limits the possible address space to 2\textsuperscript{32} (4.3 billion) addresses. The number of usable IPs is actually only 3.7 billion, since some IP ranges are designated as special-use or private \cite{ianaspprp2017}. Since not all IP addresses have a reverse DNS hostname, we parsed the Rapid7 dataset to find the actual coverage. We found that 1.25 billion addresses have a reverse DNS hostname. This finding shows that although they have significant coverage, these hostnames need to be augmented with other data to obtain a complete geolocation database.

We then quantified how many of the hostnames are valid, since the DNS records are unrestricted strings. We parsed each hostname and rejected the ones that did not respect Internet host naming rules \cite{rfc1123}. We also rejected hostnames that did not have a valid suffix as defined by the \textit{Public Suffix List}, which is a list of valid domain suffixes from the Mozilla Foundation previously described in Section \ref{sec:Datasets}. This left us with 1.24 billion hostnames, of which 1.15 billion were distinct. Our findings are summarized in Table \ref{table:Usage of Reverse DNS hostnames}, which shows that 33.4\% of usable IP addresses have a valid reverse DNS hostname, and 31.1\% are distinct. Considering that not all IPv4 addresses are yet allocated, the actual percentage is likely higher.

Next, we set out to determine if reverse DNS hostnames are a valuable source of geolocation information. We searched for exact city names and airport codes in the hostnames, using the \textit{Cities 1000} and the \textit{Alternate Names} dataset, respectively. We find that 163.7 million  hostnames could contain exact city names, and 272.9 million hostnames could contain airport codes. This approach represents an upper-bound of the number of hostnames that could contain exact city names or  airport codes. The results contain true positives such as \texttt{sur01.seattle.wa.seattle.comcast.net} in \textit{Seattle, Washington} and \texttt{inovea5.gs.par.ivision.fr} in \textit{Paris, France}. However, this naive approach also matches false negatives such as \texttt{node-j.pool-1-0.dynamic.totbb.net} which is not in \textit{Pool, UK} and \texttt{mobile.bigredgroup.net.au} which is not in \textit{Mobile, Alabama}. Nevertheless, the results summarized in Table \ref{table:Usage of Reverse DNS hostnames} show that there are potentially hundreds of millions of hostnames that could contain geographic information, using just these two features alone. We conclude that while the results are promising, a more sophisticated approach could achieve higher coverage and accuracy.

\begin{table}[t]
\caption{Usage of Reverse DNS hostnames across IPv4 space}
\label{table:Usage of Reverse DNS hostnames}
\footnotesize
	\begin{tabularx}{\columnwidth}{ @{\extracolsep{\fill}} l c c c }
		\toprule
		\textbf{Set name} & \textbf{Size} & \textbf{\% of usable} & \textbf{\% of distinct} \\
		\midrule
		\textbf{Total IPv4 space} & \textbf{4.3 B} & & \\
		\-\ \rightanglearrow[scale=1] Reserved IP addresses & 0.6 B & & \\
		\-\ \rightanglearrow[scale=1]  Usable IP addresses & 3.7 B & & \\
		\-\ \-\ \-\ \-\ \-\ \-\ \rightanglearrow[scale=1] IPs with Reverse DNS hostnames & 1.25 B & 33.7\% & \\
		\-\ \-\ \-\ \-\ \-\ \-\ \-\ \-\ \-\ \-\ \-\ \rightanglearrow[scale=1] Valid Reverse DNS hostnames & 1.24 B & 33.4\% & \\
		\-\ \-\ \-\ \-\ \-\ \-\ \-\ \-\ \-\ \-\ \-\ \-\ \-\ \-\ \-\ \-\ \rightanglearrow[scale=1] Distinct DNS hostnames & 1.15 B & 31.1\% & \\
		\-\ \-\ \-\ \-\ \-\ \-\ \-\ \-\ \-\ \-\ \-\ \-\ \-\ \-\ \-\ \-\ \-\ \-\ \-\ \-\ \-\ \-\ \rightanglearrow[scale=1] Exact city match (naive) & 0.16 B & 4.4\% & 14.1\% \\
		\-\ \-\ \-\ \-\ \-\ \-\ \-\ \-\ \-\ \-\ \-\ \-\ \-\ \-\ \-\ \-\ \-\ \-\ \-\ \-\ \-\ \-\ \rightanglearrow[scale=1] Airport code match (naive) & 0.27 B & 7.4\% & 23.5\% \\
		\bottomrule
	\end{tabularx}
\end{table}

\begin{table*}[ht!]
\caption{Top hostname components of the largest 10 domains that have reverse DNS hostnames. We manually highlighted locations with underlined blue. The percentages in the \textit{valid} and \textit{usable} columns are based on rows 3 and 5 of Table \ref{table:Usage of Reverse DNS hostnames}.}
\label{table:Top hostname n-grams}
\footnotesize
  \begin{tabularx}{\linewidth}{ llcX }
    \toprule
    \textbf{Domain} & \textbf{Count} & \textbf{\% of valid} & \textbf{Top hostname components sorted in descending order by how often they appear in all hostnames of this domain} \\
    \midrule
    
    comcast.net      & 50.0M & 4.0\% & c, hsd, hsd1, m, \boldunderofficeblue{ca}, \boldunderofficeblue{pa}, \boldunderofficeblue{fl}, \boldunderofficeblue{il}, \boldunderofficeblue{ma}, \boldunderofficeblue{ga}, a, \boldunderofficeblue{co}, \boldunderofficeblue{mi}, d, f, \boldunderofficeblue{wa}, b, e, \boldunderofficeblue{va}, \boldunderofficeblue{nj}, \boldunderofficeblue{or}, \boldunderofficeblue{md}, \boldunderofficeblue{tx}, \boldunderofficeblue{chlm}, \boldunderofficeblue{chic}, \boldunderofficeblue{phil}, \boldunderofficeblue{tn}, \boldunderofficeblue{in}, npls, dd, \boldunderofficeblue{atlt}, \boldunderofficeblue{sjos}, \boldunderofficeblue{denv}, \boldunderofficeblue{mn} \cmmnt{, , \boldunderofficeblue{bvrt}} \cmmnt{, detr} \\
    
    bbtec.net        & 37.2M & 3.0\% & softbank, biz \\
    
    rr.com           & 31.1M & 2.5\% & res, cpe, mta, \boldunderofficeblue{socal}, biz, rrcs, \boldunderofficeblue{nyc}, neo, \boldunderofficeblue{nc}, \boldunderofficeblue{wi}, kya, \boldunderofficeblue{columbus}, \boldunderofficeblue{cinci}, \boldunderofficeblue{carolina}, \boldunderofficeblue{tx}, \boldunderofficeblue{central}, \boldunderofficeblue{twcny}, \boldunderofficeblue{nycap}, \boldunderofficeblue{west}, \boldunderofficeblue{sw}, \boldunderofficeblue{rochester} \cmmnt{, \boldunderofficeblue{maine}, \boldunderofficeblue{sc}, triad} \\
    
    myvzw.com        & 29.6M & 2.4\% & sub, qarestr \\
    
    sbcglobal.net    & 28.4M & 2.3\% & lightspeed, adsl, dsl, \boldunderofficeblue{irvnca}, \boldunderofficeblue{hstntx}, \boldunderofficeblue{rcsntx}, \boldunderofficeblue{cicril}, \boldunderofficeblue{sntcca}, \boldunderofficeblue{tukrga}, \boldunderofficeblue{miamfl}, \boldunderofficeblue{pltn}, \boldunderofficeblue{pltn13}, \boldunderofficeblue{stlsmo}, \boldunderofficeblue{livnmi}, \boldunderofficeblue{bcvloh}, \boldunderofficeblue{frokca}, \boldunderofficeblue{chcgil} \cmmnt{, \boldunderofficeblue{sndgca} , snantx} \\
    
    t-ipconnect.de   & 24.5M & 2.0\% & dip, dip0, p, b, e, a, f, d, c, pd, fc, fd, fe, ff, de, dd, dc, df, ee, bb, bd, bc, ae, ac, aa, ab, af, ad, ba, bf, ea, eb, be, ec, fa, ed, fb, ef, db, da, ca, cf \cmmnt{, cc, ce} \\
    
    telecomitalia.it & 19.4M & 1.6\% & host, static, business, b, r, retail, dynamic, host156, host15, host94, host61, host127, host232, host112, host95, host72, host107, host220 \cmmnt{, host105} \\
    
    ge.com           & 16.7M & 1.4\% & static, n, n003, n003-000-000-000, n129, n144, n144-220-000-000, n129-201-000-000, n129-202-000-000, n165-156-000-000m n165, n192 \cmmnt{, n209} \\
    
    ocn.ne.jp        & 16.2M & 1.3\% & p, ipngn, \boldunderofficeblue{tokyo}, \boldunderofficeblue{osaka}, ipbf, \boldunderofficeblue{marunouchi}, ipbfp, omed, omed01, \boldunderofficeblue{kanagawa}, \boldunderofficeblue{hodogaya}, \cmmnt{,ipad,} \boldunderofficeblue{aichi}, \boldunderofficeblue{osakachuo}, \boldunderofficeblue{saitama}, \boldunderofficeblue{hokkaido} \cmmnt{, souka} \\
    
    spcsdns.net      & 16.0M & 1.3\% & pools, static \\

\bottomrule
  \end{tabularx}
\end{table*}

To further familiarize ourselves with hostname naming conventions, we extracted the top hostname components of the largest 10 domains in the Rapid7 dataset. We divided each subdomain on the dotted terms, and then we further split the components on dashes and on the transitions between numbers and letters. For example, we split \texttt{soc-l.wht2.ocn.ne.jp} into \texttt{soc}, \texttt{l}, \texttt{wht}. We then manually labeled the components that we found to reasonably correspond to geographic locations. We also cross-checked our findings with commercial geolocation databases. Table \ref{table:Top hostname n-grams} shows a sumary of the results. We observe that only 4 out of the top 10 domains contain indicators of geographic location. However, those that use geographic encodings do so extensively. We found that service providers use various naming conventions across different networks and within a single network. For instance, the hostnames under the \textit{sbcglobal.net} domain owned by \textit{AT\&T} make use of abbreviations such as \texttt{pltn} to refer to \textit{Pleasanton, CA}. But they also use combinations of city abbreviations with State names such as \texttt{chcgil} to refer to \textit{Chicago, Illinois}. Our findings are in line with previous work by Chabarek and Barford \cite{chabarek2013s}. They found that all 8 of the providers they studied used multiple naming schemes. They also found that 20 out of 22 North American providers they surveyed use geographic encodings in their hostnames.

We also studied the distribution of top-level domains (TLDs) such as \texttt{.com} and \texttt{.fr} in the \textit{Rapid7} dataset to determine if country-specific domains can be used as location hints. We observed that most hostnames contain a \texttt{.net} domain at 33.2\%, followed by \texttt{.com} with only 17.2\%. This is the opposite of forward DNS, where \texttt{.com} is more popular. The difference is due to Internet Service Providers preferring to use \texttt{.net} domains for hostnames that describe the underlying physical architecture of their \textit{network}. After removing the \texttt{.com}, \texttt{.net}, \texttt{.edu}, and \texttt{.mil} domains which together make up 51.6\% of valid hostnames, we are left with approximately 600 million hostnames, the vast majority of which are country-specific. We found very few novelty TLDs used in reverse DNS hostnames. We conclude that the corresponding country of a reverse DNS domain could potentially be a useful hint in geolocation.

Finally, we compared snapshots of the dataset, each collected in the month of January of years 2014 to 2017, inclusive. Our goal was to determine how the characteristics of the hostnames change in time. For each IP in the snapshot we compared the hostname values in consecutive years. Table \ref{table:Reverse DNS hostname changes across 4 years} shows a summary of the results. We found that a maximum of 14.7\% of hostnames changed year over year and 63.7\% of them remain the same across all four years. These numbers include the cases where one side of the comparison had a hostname but the other side was empty due to the DNS query returning an empty hostname, or due to the request failing because of network failures during data collection. We then performed a similar comparison, this time counting only the cases where both sides of the comparison contained non-empty hostnames. Here we found that a maximum of 2.2\% hostnames change over the years, if both the values are present. To understand why there is such a large discrepancy between these two findings we also determined the number of hosts that were gained or list between the years. By hostnames gained we mean that in the older year a hostname was missing, while in the subsequent year it was present, and by hostnames lost we mean the opposite. The results show that yearly more hostnames are gained than lost, at about a ratio of 2:1. 

In summary, we determined that 1.24 billion IP addresses have valid reverse DNS hostnames with 1.15 billion distinct values, many of which contain exact city or airport code matches.

\begin{table}[t]
\caption{Reverse DNS hostname changes across 4 years}
\label{table:Reverse DNS hostname changes across 4 years}
\footnotesize
	\begin{tabularx}{\columnwidth}{ @{\extracolsep{\fill}} l c c c }
		\toprule
		\textbf{Change / Year Pair} & \scriptsize{\textbf{2014 \textrightarrow 2015}} & \scriptsize{\textbf{2015 \textrightarrow 2016}} & \scriptsize{\textbf{2016 \textrightarrow 2017}} \\
		\midrule
		Hostnames changed (incl. empty) &  \scriptsize{\textbf{179M} (14.7\%)} & \scriptsize{\textbf{159.3M} (12.6\%)} & \scriptsize{\textbf{164.8M} (12.8\%)} \\
		Hostnames changed (non-empty) & \scriptsize{\textbf{26.4M} (2.2\%)} & \scriptsize{\textbf{20.7M} (1.6\%)} & \scriptsize{\textbf{14.2M} (1.1\%)} \\
		Hostnames gained & \scriptsize{\textbf{108.5M} (8.9\%)} & \scriptsize{\textbf{89.3M} (7.1\%)} & \scriptsize{\textbf{81M} (6.3\%)} \\
        Hostnames lost & \scriptsize{\textbf{44M} (3.6\%)} & \scriptsize{\textbf{49.4M} (3.9\%)} & \scriptsize{\textbf{70M} (5.4\%)} \\
		\bottomrule
	\end{tabularx}
\end{table}








\section{Approach}
\label{sec:Approach}

We cast the problem of extracting locations from reverse DNS hostnames as a machine learning problem. We train a binary classifier on a dataset where each training sample is a hostname and location candidate pair, along with a binary label which signifies if the hostname is \textit{likely} or \textit{unlikely} to be in the candidate location. Given a new hostname, our proposed approach splits the hostname into components, finds a preliminary list of location candidates, generates primary and secondary features for each candidate, then classifies each potential location using the classifier, also assigning each candidate a confidence score. For instance, for the hostname \texttt{ce-salmor0w03w.cpe.or.portland.bigisp.net} our approach considers tens of potential location candidates, including \textit{Portland, UK} and \textit{Salmoral, Spain}. In the end however, it ranks \textit{Salem, Oregon} and \textit{Portland, Oregon} as the most likely candidates.


\subsection{Splitting hostnames}
\label{sec:Splitting hostnames}


Drawing from our preliminary analysis in Section \ref{sec:Reverse DNS}, as well as further manual analysis, we implemented multiple heuristics for splitting hostnames into their constituent components. 

First, we apply the \textit{ToUnicode} algorithm described in RFC3490 \cite{rfc3490} to convert International Domain Names (IDN) to Unicode. The reason we perform this translation is that international hostnames are stored as ASCII strings using Punycode transcription. For example, the hostname \texttt{xn--0rsod70av79j.xn--j6w193g} gets converted to \begin{CJK*}{UTF8}{gbsn}夏威夷舞.香港\end{CJK*}. This allows us to perform location lookups using the original language of the hostname.

Second, we separate the subdomain from the domain and the public suffix, using the list provided by the Mozilla Foundation previously described in Section \ref{sec:Datasets}. These suffixes are a superset of normal TLDs because they also contain entire domains under which users can create subdomains. For example, the list contains the pseudo-TLD \texttt{azurewebsites.net} since users of Azure cloud services can register their own subdomains under this name. At this point we also extract the native TLD. For instance, for \texttt{dps8099.denver.k12.co.us} we extract \texttt{denver.k12.co.us} as the domain because \texttt{k12.co.us} is a public suffix, we extract \texttt{dps8099} as the subdomain, and finally we extract \texttt{.us} as the TLD.

\begin{figure}
\centering
\includegraphics[width=0.7\columnwidth]{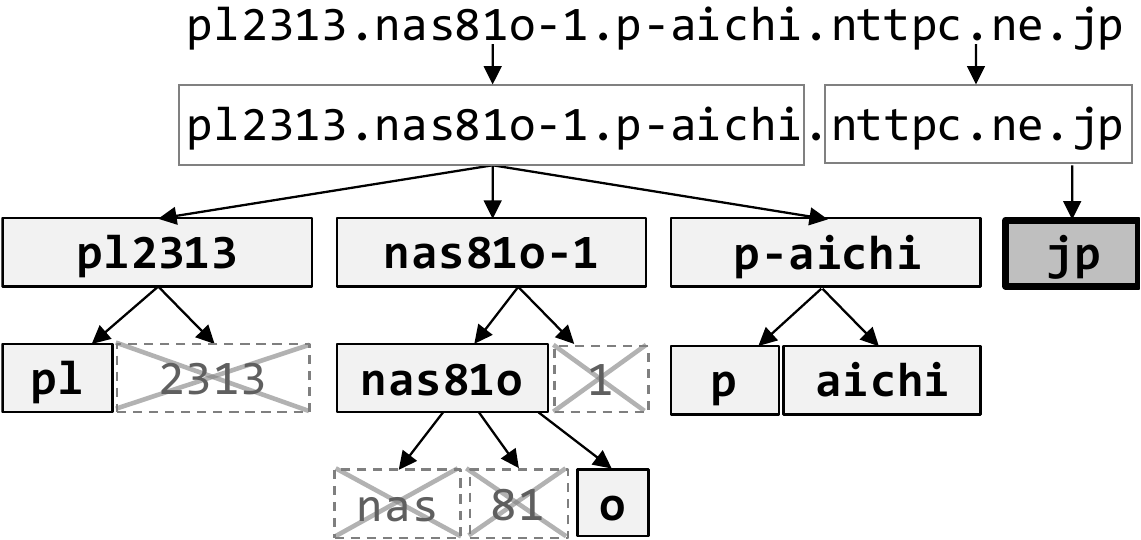}
\vspace{-.1in}
\caption{Hostname Splitter example with pruning}
\label{fig:HostnameSplitter}
\end{figure}

Third, we split the extracted subdomain at three levels of aggregation: on the dotted elements, on hyphens within the dotted elements, and on the transitions between letters and numbers within the hyphenated elements, saving the results at each level. Figure \ref{fig:HostnameSplitter} contains a specific example represented intuitively as a tree structure. The bottom three levels of the tree correspond to the three levels of aggregation. As a last step, we trim the leaf nodes. We remove any leaf node consisting solely of numbers. We also remove common terms terms related to connection characteristics, such as \texttt{dsl}, \texttt{fiber}, and \texttt{nas}. We obtained them by counting the top extracted leaf nodes in the training set and manually selecting the ones which are unrelated to geolocation but clearly related to the underlying network infrastructure. The list is available in the source code we are publishing along with this paper.

\subsection{Features}
\label{sec:Features}

\begin{figure}
\centering
\includegraphics[width=1\columnwidth]{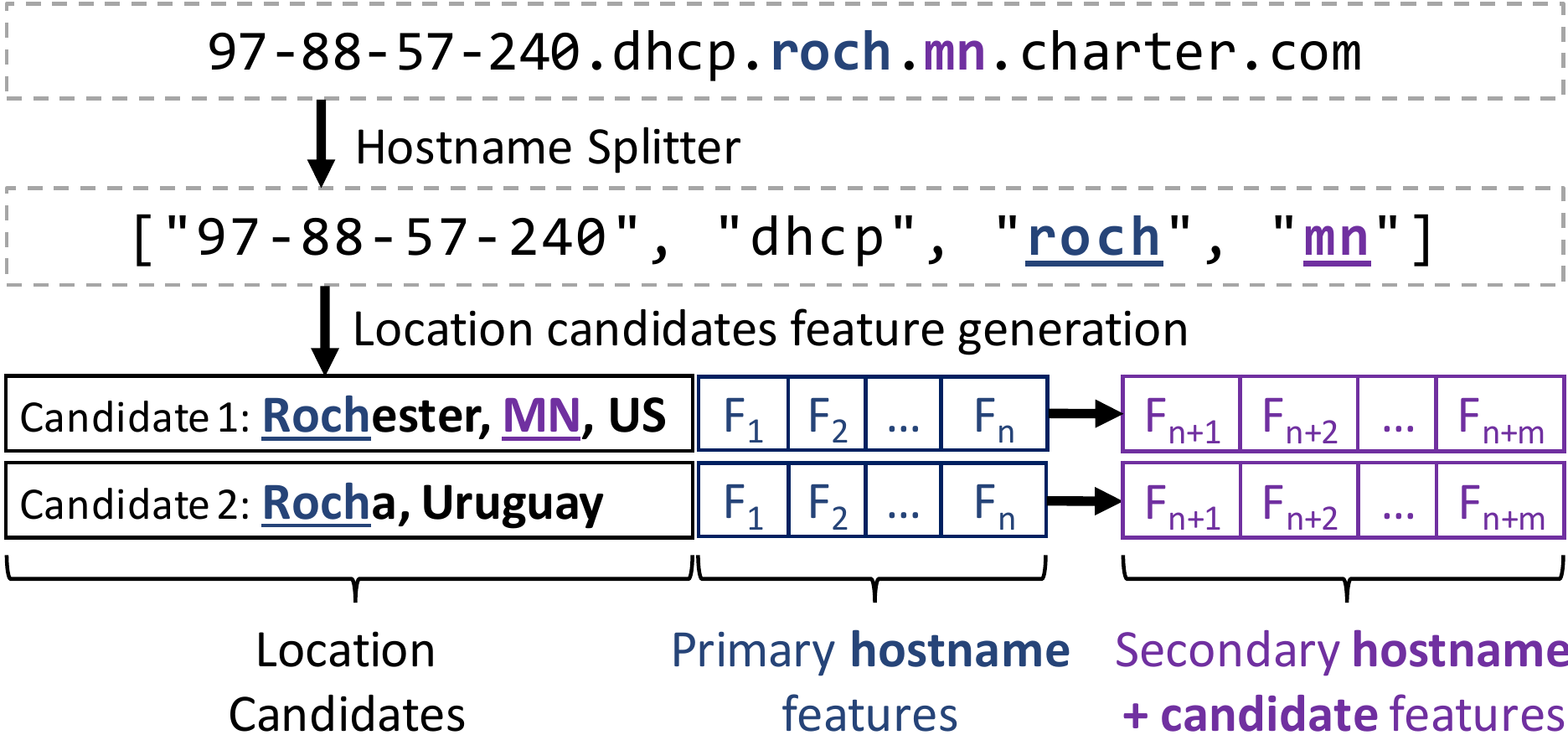}
\vspace{-.1in}
\caption{Feature Matching and Generation}
\vspace{-.1in}
\label{fig:FeatureGeneration}
\end{figure}

Starting from the results of the hostname splitter, we find the location candidates along with their {\em primary} and {\em secondary} features, as defined below. \textbf{The list of preliminary location candidates is defined by the union of locations which match any of the primary features of the hostname.} Figure \ref{fig:FeatureGeneration} shows a concrete example. Primary features can be derived directly from a hostname. These features are matched using a single contiguous string which indicate a location at city level granularity. Primary feature generation and candidate selection happen at the same time. Secondary features are generated in the context of a hostname and location candidate pair. These features require the context of a primary candidate to match. In our example two location candidates and their primary features are first selected based on the term \texttt{roch} in the hostname. Then we compute secondary features separately for each candidate. In the context of \textit{Rochester, Minnesota}, we match the \texttt{mn} term as a secondary feature that captures the administrative region for this candidate.

\textbf{Primary features} are based on the \textit{GeoNames}, \textit{UN/LOCODE}, and \textit{CLLI} datasets described in Section \ref{sec:Datasets}. From \textit{GeoNames} we use the \textit{Cities 1000}, \textit{Alternate Names}, and \textit{Admin 1 Codes} subsets. The primary feature categories are listed in Table \ref{table:Primary Features Categories}. Each of these categories is represented by three specific features: \textit{IsMatch}, \textit{Population}, and \textit{MatchedLettersCount}. The \textit{IsMatch} feature is a boolean which indicates if the feature matched the current hostname and current location candidate. The \textit{Population} feature contains the population of the current location, if \textit{IsMatch} is \textit{true}. We use population as a proxy for the importance of a city candidate. Finally, \textit{MatchedLettersCount} contains the number of characters which matched. As the number of characters in common between a hostname and a location increases, it could mean a higher confidence match. For instance, if the hostname contains the letters \texttt{seattle} and the current location candidate is \textit{Seattle, Washington}, then the \textit{CityName-MatchedLettersCount} Feature would have a value of seven.

\begin{table}[t]
\caption{Primary Feature Categories}
\vspace{-.1in}
\label{table:Primary Features Categories}
\footnotesize
  \begin{tabularx}{\columnwidth}{ lll }
    \toprule
    \textbf{Category} & \textbf{Example} & \textbf{Location} \\
    \midrule
    City Name & p907072-li-mobac01.{\boldunderofficeblue{osaka}}.ocn.ne.jp & Osaka, JP \\
    Alternate names & 178235248188.{\boldunderofficeblue{warszawa}}.vectranet.pl & Warsaw, PL \\
    Abbreviations & cpe-68-173-83-248.{\boldunderofficeblue{nyc}}.res.rr.com & New York City \\ 
    City + Admin1 & {\boldunderofficeblue{torontoon}}-rta-1.inhouse.compuserve.com & Toronto, ON \\
    City + Country & er1-ge-7-1.{\boldunderofficeblue{londonuk}}5.savvis.net & London, UK \\
    No Vowels Name & static-50-47-60-130.{\boldunderofficeblue{sttl}}.wa.frontiernet.net & Seattle, WA \\
    First Letters & 97-90-205-107.dhcp.{\boldunderofficeblue{losa}}.ca.charter.com & Los Angeles \\
    Airport Code & 62.80.122.50.{\boldunderofficeblue{fra}}.de.eunx.net & Frankfurt, DE \\
    CLLI Code & 99-166-111-251.{\boldunderofficeblue{tukrga}}.sbcglobal.net & Tucker, GA \\
    UN/LOCODE & 16.151.88.129,{\boldunderofficeblue{krsel}}19d.kor.hp.com & Korea, Seoul \\
    Host Patterns & {\boldunderofficeblue{atoulon}}-651-1-29-109.abo.wanadoo.fr & Toulon, FR \\

    \bottomrule
  \end{tabularx}
\end{table}

\begin{table}[t]
\caption{Secondary Features Categories}
\vspace{-.1in}
\label{table:Secondary Features Categories}
\footnotesize
  \begin{tabularx}{\columnwidth}{ lll }
    \toprule
    \textbf{Category} & \textbf{Candidate} & \textbf{Match Example} \\
    \midrule
    Admin1 & Johnstown, PA & 138-207-246-119.{\boldunderofficeblue{jst}}.{\boldunderofficeviolet{pa}}.atlanticbb.net \\
    First Letters Admin 1 & Ft. Huachuca, AZ & {\boldunderofficeblue{frth}}-bw-noc.{\boldunderofficeviolet{ariz}}.aisco.ngb.army.mil \\
    Country & Paris, FR & ci77.{\boldunderofficeblue{paris}}12eme.{\boldunderofficeviolet{fr}}.psi.net \\
    Country TLD & Barcelona, ES & {\boldunderofficeblue{barcelona}}.fib.upc.{\boldunderofficeviolet{es}} \\
    \bottomrule
  \end{tabularx}
  \vspace{-.1in}
\end{table}

While most feature categories in Table \ref{table:Primary Features Categories} are self-explanatory, we describe them here briefly. The \textit{City Name} category matches entire names of cities. \textit{Alternate names} matches translations and colloquial names of locations. \textit{Abbreviations} are based on the first letters of cities with longer names, such as \texttt{sf} for \textit{San Francisco}. The \textit{City + Admin1} category consists of concatenations of city and administrative regions, such as \texttt{seattlewa}. Similarly, \textit{City + Country} matches combinations of city and country names. 

The intent of the \textit{No Vowel Name} feature is to match city names without vowels. It allows partial matches using the first 3 or more letters of the names. For example, this allows matching \texttt{gnvl} to \textit{Greenville, SC} and \texttt{rvrs} to \textit{Riverside, CA}. Furthermore, based on our observations we extended this feature with more complex variations. We select the first and last letters of each word in the name, even if the letters are vowels. We then generate combinations of letters from this list, in order. Examples matched by this variation include \texttt{oxfr} for \textit{Oxford, MA}, and \texttt{ftmy} for \textit{Fort Meyers, FL}.

The \textit{First Letters} features use the first consecutive letters of locations. The \textit{Airport Code} category spans airport codes from travel organizations. \textit{CLLI} and \textit{UN/LOCODE} codes match telecommunications and transportation codes of locations, respectively. 


Finally, \textit{Host Patterns} attempts to capture rules not encompassed by the other features. Using training data we extract frequently co-occurring hostname term permutations of one or two terms. We then aggregate the training data per domain and within a domain on the term permutations. If at least 40\% of the training locations for a permutation are located within a 20 kilometer radius, we convert the term permutation into a rule. We determined the support ratio and the distance radius using a small validation set. For example, this feature determines that whenever a hostname in the \texttt{frontiernet.net} domain contains the term \texttt{or} in the rightmost position and the term \texttt{mmvl} in the second rightmost position, then the hostname is most likely located in \texttt{McMinnville, Oregon}. This feature would then match for the hostname \texttt{static-50-126-80-6.mmvl.or.frontiernet.net}.

\textbf{Secondary features} are determined in the context of a hostname and location candidate pair. As shown in Figure \ref{fig:FeatureGeneration}, we first determine all candidates before we can compute the secondary features. An example of secondary features for the \textit{Rochester, MN} candidate is \textit{Admin1 Match}, which is \textit{true} only if the administrative region of the candidate location  can be found in a different term of the hostname. Since the hostname contains the term \texttt{mn}, which is an abbreviation of Minnesota, then this secondary feature is \textit{true} for the first candidate. However, it is \textit{false} for the second candidate, because \textit{Rocha} is in an administrative region also called \textit{Rocha}, and it cannot be found in the hostname. \textit{First Letters Admin 1} is similar, but it matches at least 3 first consecutive letters of administrative names. \textit{Country} and \textit{Country TLD} both try to match the country of the current candidate by searching for a country code in the hostname terms or in the domain \textit{TLD}, respectively. 



\subsection{Classifier}
\label{sec:Classifier}

For a given hostname, our reverse DNS geolocation can extract and evaluate tens of potential location candidates. For example, if one of the terms of the hostname is \texttt{york}, the initial list of candidates will contain all locations named \textit{York} in the world. We run a binary classifier on each of the initial candidates. The classifier uses the primary and secondary features to evaluate if it is plausible for the hostname to be located in a candidate location. All the candidates where the classifier returns \texttt{false} are discarded. The remaining plausible candidates are sorted by confidence and returned in a list.


Although determining the optimal type of binary classifier is outside of the scope of this work, we tested four variations of the classifier: logistic regression, C4.5 decision trees, random forest, and SVM. Logistic regression had the best performance on a small validation set. Consequently, we performed all experiments in Section \ref{sec:Evaluation} using this classifier.

\subsection{Sampling Strategy}
\label{sec:Sampling Strategy}

We propose sampling the training set to account for data bias, to improve generalization, and to reduce the amount of required training data. First, the entire set of reverse DNS hostnames is naturally skewed towards the largest Internet Service Providers, which own the most addresses. Second, some feature categories such as \textit{City Name} occur much more often than others such as \textit{Abbreviations}. This can lead the classifier to ignore less frequent features categories. Third, during training multiple location candidates can be generated for each hostname, out of which at most one can be correct. Since the classifier is trained on hostname and candidate pairs, this also introduces another type of bias where the number of negative samples significantly outweighs the number of positive ones. Therefore, we sample data to account for some of this bias and to improve generalization through increased training data diversity.

We perform stratified sampling on the domain of the hostname, keeping at most $\mathcal{X}$ samples per domain. This approach ensures that naming schemes of large organizations do not significantly skew the training data. We further increase feature diversity by keeping a ratio of $\mathcal{Y}:1$ between the number of samples that contain the most commonly occurring feature and the ones that contain the least occurring feature. Finally, we also enforce a ratio of $\mathcal{Z}:1$ between the number of negative and positive examples. We evaluate our data sampling strategy and its three parameters in Section \ref{sec:Preliminary Evaluation}.

\section{Evaluation}
\label{sec:Evaluation}

We evaluate our approach against three state-of-the-art academic baselines and two commercial geolocation databases. We show that our method significantly outperforms academic baselines and is complementary and competitive to commercial databases.

\subsection{Ground Truth}
\label{sec:Ground Truth}

Our ground truth dataset contains 67 million IP addresses with known IP location, of which we used 40 million for training and 27 million for testing. We compiled the dataset in March 2018 from a subset of the query log of a major search engine. Each IP address has a corresponding location obtained from users that opted in to provide their location through devices connected to cellular networks or home Wi-Fi networks. We discarded any IP address that was present in multiple cities over the course of a month. The locations were aggregated at IP and city level by an automated pipeline. We did not have access to the locations of individual users.

\subsection{Preliminary Evaluation}
\label{sec:Preliminary Evaluation}

We conducted two experiments to evaluate the binary classifier in isolation. In the first experiment, we randomly selected 100,000 IP addresses from the training set and performed ten-fold cross validation. We did not further sample the data in any other way. For each hostname, we extracted location candidates, then ran the binary classifier on all the pairs between the target hostname and each of its candidates. Since \textbf{our approach can return multiple plausible locations for a given hostname}, we  choose the candidate with the highest classifier confidence. We break ties by selecting the location with the highest population, as a proxy of popularity. We obtained an overall accuracy of 99\%, mostly because the vast majority of results were true negatives. However, the true positive rate was only 67.6\%, precision was 80.9\%, and recall was 67.6\%.

In the second experiment we introduced training data sampling as described in Section \ref{sec:Sampling Strategy}. We set the $\mathcal{X}$, $\mathcal{Y}$, and $\mathcal{Z}$ parameters to \textit{200}, \textit{10}, and \textit{3}, respectively. We again performed ten-fold cross validation. Although accuracy decreased to 92.9\%, we obtained better results for true positive rate, precision, and recall, at 78.8\%, 88.5\%, and 78.8\%, respectively. We varied the values of the $\mathcal{X}$, $\mathcal{Y}$, and $\mathcal{Z}$ parameters using exhaustive search but this did not alter the results significantly. In conclusion, our sampling strategy helps the classifier generalize and it significantly improves results. 


\subsection{Academic baselines}
\label{sec:Academic baselines}

\begin{table*}[t]
\caption{Evaluation against three state of the art academic baselines. \textit{undns} from University of Washington consists of manual rules, \textit{DRoP} from University of California's CAIDA uses automatically generated rules, and \textit{DDec} also from CAIDA uses a combination of the first two. Results for our approach, which is fully automated, are under the \textit{RDNS} heading.}
\vspace{-.1in}
\label{table:Academic Evaluation}
\scriptsize
  \begin{tabularx}{\textwidth}{ lr|rrrr|rrrr|rrrr|rrrr }
    \toprule
    \multicolumn{2}{l|}{\textbf{Metric} \hspace{0.4mm} \textrightarrow } & \multicolumn{4}{c|}{\textbf{Median Error in km} (lower is better)} & \multicolumn{4}{c|}{\textbf{RMSE based on km} (lower is better)} & \multicolumn{4}{c|}{\textbf{Coverage} (higher is better)} & \multicolumn{4}{c}{\textbf{Combined score} (higher is better)} \\
    \midrule
    \textbf{Domain} \textdownarrow & \# & \textbf{undns} & \textbf{DRoP} & \textbf{DDec} & \textbf{RDNS} & \textbf{undns} & \textbf{DRoP} & \textbf{DDec} & \textbf{RDNS} & \textbf{undns} & \textbf{DRoP} & \textbf{DDec} & \textbf{RDNS} & \textbf{undns} & \textbf{DRoP} & \textbf{DDec} & \textbf{RDNS} \\

163data.com.cn       & 166K & 1,517.5 & N/A      & 1,517.5 & \boldunder{10.6} & 1,495 & N/A      & 1,495 & \boldunder{404}   & \boldunder{100\%} & N/A    & \boldunder{100\%} & 94.5\% & 0.67 & N/A  & 0.67 & \boldunder{2.34} \\
bell.ca              & 200K & N/A     & 5,875.2  & 5,875.2 & \boldunder{6.0}  & N/A     & 5,807  & 5,807 & \boldunder{1,262} & N/A     & 2.3\%  & 2.3\%   & \boldunder{95.7\%} & N/A  & 0.00 & 0.00 & \boldunder{0.76} \\
brasiltelecom.net.br & 32K & 808.7   & 5,628.7  & 808.7   & \boldunder{15.2} & 889   & 5,620  & 889   & \boldunder{427}   & \boldunder{100\%} & 69.7\% & \boldunder{100\%} & 73.9\% & 1.12 & 0.12 & 1.12 & \boldunder{1.73} \\
charter.com          & 580K & 60.8    & N/A      & 60.8    & \boldunder{59.9} & \boldunder{478}   & N/A      & \boldunder{478}   & 484   & 78.0\%  & N/A  & 78.0\%  & \boldunder{89.0\%} & 1.63 & N/A  & 1.63 & \boldunder{1.84} \\
frontiernet.net      & 67K & 36.5    & 6,247.6  & 36.5    & \boldunder{16.7} & 785   & 6,101  & 785   & \boldunder{689}   & 3.6\%   & 0.8\%  & 3.6\%   & \boldunder{99.4\%} & 0.05 & 0.00 & 0.05 & \boldunder{1.44} \\
nttpc.ne.jp          & 0.9K & 9.5     & 9,259.9  & 16.2    & \boldunder{9.1}  & \boldunder{2,081} & 9,161  & 4,976 & 3,694 & 12.0\%  & 16.2\% & 16.2\%  & \boldunder{57.6\%} & 0.06 & 0.02 & 0.03 & \boldunder{0.16} \\
optusnet.com.au      & 100K & 704.4   & 16,134.6 & 704.4   & \boldunder{12.7} & 1,175 & 16,374 & 1,175 & \boldunder{583}   & \boldunder{100\%} & 49.8\% & \boldunder{100\%} & 98.9\% & 0.85 & 0.03 & 0.85 & \boldunder{1.70} \\
qwest.net            & 408K & 3,426.6 & 8,038.7  & 8,038.7 & \boldunder{17.6} & 6,856 & 7,361  & 7,361 & \boldunder{427}   & 0.0\%   & 4.1\%  & 4.1\%   & \boldunder{94.0\%} & 0.00 & 0.01 & 0.01 & \boldunder{2.20} \\

	\midrule

Overall              & 1.6M & 163.9   & 13,974.2 & 177.9   & \boldunder{17.5} & 924.0   & 12,640.4 & 1,497.5 & \boldunder{677.8}   & 48.3\%  & 6.1\%  & 49.7\%  & \boldunder{92.3\%} & 0.52 & 0.00 & 0.33 & \boldunder{1.36} \\
    
    \bottomrule
    
  \end{tabularx}
\end{table*}

\begin{table*}[t]
\caption{Examples of locations extracted incorrectly by the \textit{DRoP} \textbf{baseline}}
\vspace{-.1in}
\label{table:Incorrect DRoP Examples}
\scriptsize
  \begin{tabularx}{\textwidth}{ llll }
    \toprule
    \textbf{Hostname} & \textbf{Location extracted incorrectly by \textit{DRoP}} & \textbf{Correct location} & \textbf{\textit{DRoP} Rule} \\
    \midrule
    \texttt{\boldunder{malton}2259w-lp140-03-50-100-186-228.dsl.bell.ca} &  \textbf{malton} → Malton, North Yorkshire, England & \textbf{malton}, \textbf{.ca} → Malton, Canada & \texttt{\%\verb|<<pop>>([^L]+L+D*){3}.bell.ca|} \\    
    \texttt{200-96-182-198.cbace700.\boldunder{dsl}.brasiltelecom.net.br} & \textbf{dsl} → Daru, Sierra Leone & \textbf{cbace}, \textbf{.br} → Cuiabá, Brazil & \texttt{\%\verb|<<iata>>.brasiltelecom.net.br|} \\    
    \texttt{70-100-143-28.dsl2-\boldunder{pixley}.roch.ny.frontiernet.net} & \textbf{pixley} → Pixley, California, USA & \textbf{roch}, \textbf{ny} → Rochester, New York, USA & \texttt{\%\verb|<<pop>>([^L]+L+D*){2}.frontiernet.net|} \\    
    \texttt{st0120.\boldunder{nas}931.m-hiroshima.nttpc.ne.jp} & \textbf{nas} → Nassau, Bahamas & \textbf{hiroshima}, \textbf{.jp} → Hiroshima, Japan & \texttt{\%\verb|<<iata>>([^L]+L+D*){2}.nttpc.ne.jp|} \\ 
    \texttt{d49-194-53-51.meb1.\boldunder{vic}.optusnet.com.au} & \textbf{vic} → Vicenza, Italy & \textbf{meb}, \textbf{vic}, \textbf{.au} → Melbourne, Victoria & \texttt{\%\verb|<<iata>>.optusnet.com.au|} \\ 
    \texttt{71-209-14-48.\boldunder{bois}.qwest.net} & \textbf{bois} → 's-Hertogenbosch, The Netherlands & \textbf{bois} → Boise, Idaho, USA & \texttt{\%\verb|<<pop>>.qwest.net|} \\ 
    \bottomrule
    
  \end{tabularx}
\end{table*}

We next evaluate against three state-of-the-art academic baselines. Like our approach, they receive a hostname as input and attempt to extract its location. The \textit{undns} baseline from University of Washington consists of manually generated rules that map hostname patterns to locations \cite{undns2002}. The \textit{DRoP} baseline from CAIDA at University of California-San Diego relies on automatically generated rules derived from hostname patterns and validated by active measurement data (traceroutes) \cite{drop2014}. Finally, the \textit{DDec} baseline also from CAIDA combines the results from \textit{undns} and \textit{DRoP} \cite{ddec2018}. 

Since all three baselines are accessed from a public web endpoint \cite{ddec2018}, we had to restrict the number of requests we made to a manageable size, out of politeness. For testing we initially selected multiple service providers of different sizes, spanning various countries around the world. However, the baselines were missing \textbf{any} rules for several of these providers, including \textit{airtelbroadband.in} from India, \textit{bigpond.net.au} from Australia, and \textit{megared.net.mx} in Mexico. Although the baselines have good rule coverage in North America, they are at least partially lacking in international coverage. In the interest of fairness, we selected a list of eight providers, each of which are covered by at least two of the baselines. 

To train the classifier, our sampling strategy only considered approximately 60,000 data points out of the 40 million hostnames in our training set. From our test set of 27 million IP addresses, we selected all of the ground truth data points which intersected the eight target providers, which yielded a testing subset of 1.6 million hostnames. We issued these requests to the CAIDA web endpoint and parsed the responses from each of the baselines.

Table \ref{table:Academic Evaluation} lists the results for each of the eight domains, as well as the overall results across the entire testing subset. Our approach is labeled \textit{RDNS} in the table. We define the \textit{error distance} in kilometers to be the distance between where a model places the location of a hostname, and the actual location of the IP address behind that hostname. The first block of results shows median error distance in kilometers. We observe that \textbf{our model significantly outperforms the baselines} and its results are generally more stable across all domains. We also observe that the median error distance for several domains is abnormally high for the \textit{DRoP} baseline, and sometimes for the other baselines as well. To further investigate this surprising finding we manually verified a small sample of results. Table \ref{table:Incorrect DRoP Examples} lists examples of locations extracted incorrectly by the \textit{DRoP} baseline. In the last column of the table we list the rule that caused the incorrect extraction. For example, \textit{DRoP} incorrectly determines that the hostname \texttt{d49-194-53-51.meb1.\boldunder{vic}.optusnet.com.au} is in Vicenza, Italy, using the rule \texttt{\%<<iata>>.optusnet.com.au}. Although the \textit{IATA} airport code \textit{vic} is indeed located in Vicenza, the correct location is Melbourne, Victoria. We could not find any \texttt{optusnet.com.au} hostname where the rule was correct. In conclusion, the \textit{DRoP} baseline contains incorrect rules for some domains. The results for the \textit{undns} baseline also indicate high error distance for multiple test domains. After investigating the results, we found that \textit{undns} sometimes maps entire TLDs to a single city. For example, the locations for all \texttt{163.data.com.cn} hostnames are extracted as \textit{Beijing, China}. Lastly, since \textit{DDec} is a combination of \textit{undns} and \textit{DRoP}, it is also affected by incorrect rules.

The advantage of using median as a metric is that it is impervious to outliers, which can favor our model that can place false positives far from the actual location, generating larger  outliers. To fairly characterize the results, we also computed \textit{RMSE}, a metric at the other extreme of the spectrum. \textit{RMSE}, which stands for root mean squared error, easily gets swayed by large outliers. This poses a disadvantage for our model. We compute it using the error distance in kilometers for each hostname. The \textit{RMSE} results in Table \ref{table:Academic Evaluation}  show that generally our approach still outperforms the baselines in 6 out of 8 domains. In the two cases where our model has higher \textit{RMSE} than the models, the coverage of our model is higher.

In 3 out of 8 cases the \textit{undns} baseline has 100\% coverage. We define coverage as the total number of hostnames where a model made a decision, over the total number of hostnames in the test set. \textit{undns} having high coverage is a side effect of it using catch-all rules that map entire TLDs to a single city. In all three cases this leads to poor results for both median error and \textit{RMSE}.

We define the combined score as the inverse of \textit{RMSE} multiplied by coverage. As error distance improves (gets smaller), the combined score increases, and vice versa. Similarly, higher coverage also improves the combined score, and vice versa. \textbf{Our approach significantly outperforms all academic baselines} when considering the combination of error distance and coverage.

\begin{figure}
\centering
\includegraphics[width=1\columnwidth]{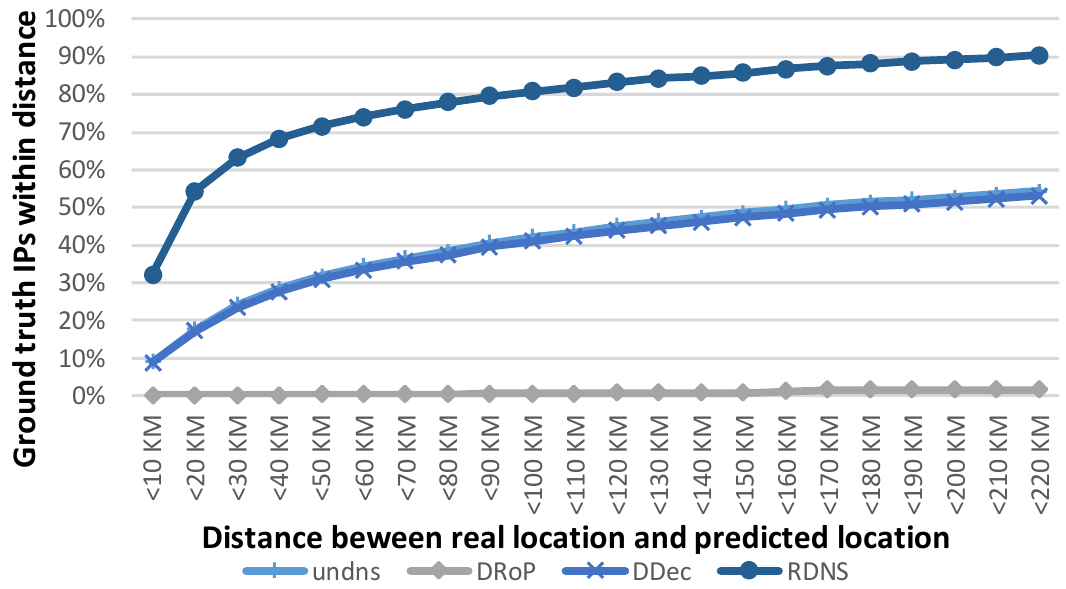}
\caption{Academic Evaluation Error Distance}
\label{fig:Academic Evaluation}
\end{figure}

Finally, Figure \ref{fig:Academic Evaluation} displays the cumulative error distance in kilometers. The X axis represents the maximum distance between the real location and the predicted location. The Y axis shows how many hostnames and their IP addresses fall within the error distance. For instance, the \textit{<20 km} column shows that our method, labeled \textit{RDNS}, places approximately 54\% of hostnames in the ground truth set within 20 kilometers of their actual location. Our method outperforms the baselines by a large margin. The \textit{DRoP} baseline yields the worst results, significantly underperforming the other methods.

\subsection{Commercial baselines}
\label{sec:Commercial baselines}

In this work we focus on improving reverse DNS geolocation, which is only one source of geolocation information. Table \ref{table:Usage of Reverse DNS hostnames} reveals that about a third of IP addresses have reverse DNS hostnames. A further subset of these hostnames contain location hints. While this can result in hundreds of millions of hostnames with location information, this is insufficient to completely cover the \textit{IPv4} space.

Commercial geolocation databases combine and conflate multiple geolocation data sources. Information from reverse DNS hostnames is required but not sufficient to compile a full geolocation database. Our approach, which can output multiple potentially valid location candidates for a given hostname, lends itself to being combined with other data source to form a more complete database.

\begin{figure}
\centering
\includegraphics[width=1\columnwidth]{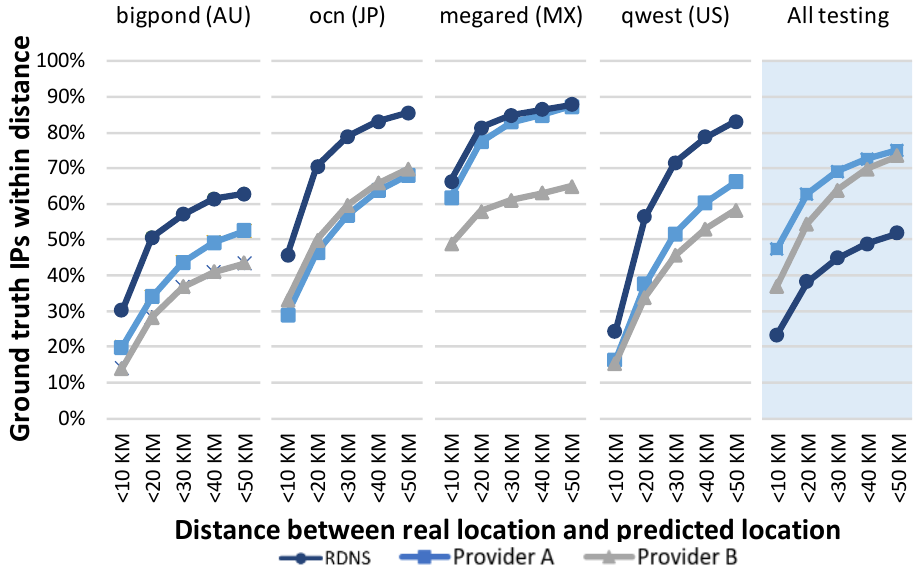}
\caption{Commercial Evaluation Error Distance}
\label{fig:Commercial Evaluation}
\end{figure}

Although reverse DNS geolocation on its own cannot match commercial databases, we evaluate our approach to show that our approach can complement and potentially improve existing databases. We trained our classifier as described in Section \ref{sec:Academic baselines}. We then obtained two state of the art commercial IP geolocation databases. We tested our approach against the two commercial database providers \textit{A} and \textit{B} using our entire test dataset of 27 million hostnames. The first four graphs in Figure \ref{fig:Commercial Evaluation} show that on certain domains our approach outperforms, and thus can be used to improve, commercial databases. However, as expected, the fifth graph shows that overall the commercial databases still outperform our method. Results show that median error is 43.7, 16.7, 11.1 kilometers, and \textit{RMSE} is 4649, 545.3, 545.9 for \textit{RDNS}, \textit{Provider A}, and \textit{Provider B}, respectively. 

\section{Privacy}
\label{sec:Privacy}


Online privacy is becoming increasingly important. For example, Pew research has found in 2016 that while many Americans are willing to share personal information in exchange for accessing online services, they are often cautious about disclosing their information and frequently unhappy about what happens to that information once companies have collected it \cite{rainie2016privacy}. We have designed both our approach and our evaluation with this sensitive subject in mind. 

Our proposed geolocation method relies on reverse DNS hostnames shared publicly by Internet Service Providers. These hostnames provide only coarse city-level or region level location information. Therefore our approach may be more privacy conscious than the widespread industry practice of requesting exact GPS coordinates through mobile apps or the HTML5 Geolocation API. 

Furthermore, the ground truth set was anonymized before we had access to it by modifying raw locations in a random direction by 200 meters, aggregating all locations reported for an IP address, and reducing location accuracy to city-level. 




\section{Reproducing Results}
\label{sec:Reproducing Results}

To aid in reproducing and extending our results, we are open sourcing all the major components of our approach, including the hostname splitter and the terms blacklist, our sampling strategy, the primary and secondary feature generators, as well as the classifier itself. For feature generation we have purposely used mainly freely available datasets as described in Section \ref{sec:Datasets}. While we cannot include our ground truth set because it is proprietary, we will make available a binary version of our model. We will also publish instructions on creating a ground truth set using public datasets by using our sampling strategy to minimize any manual labeling.

\section{Conclusions and Future Work}
\label{sec:Conclusions and Future Work}

We presented a machine learning approach to geolocating reverse DNS hostnames. Our method significantly outperforms several state of the art academic baselines and it is competitive and complementary with commercial baselines. Our method outputs multiple plausible locations in case of ambiguity. It thus lends itself to being combined with other data sources to form a more complete geolocation database. Our future work will focus on combining reverse DNS hostname information with WHOIS databases and network delay to form a geolocation database across the entire IP space.


%
\bibliographystyle{ACM-Reference-Format}
\bibliography{sigproc}  


\begin{thebibliography}{00}


\ifx \showCODEN    \undefined \def \showCODEN     #1{\unskip}     \fi
\ifx \showDOI      \undefined \def \showDOI       #1{#1}\fi
\ifx \showISBNx    \undefined \def \showISBNx     #1{\unskip}     \fi
\ifx \showISBNxiii \undefined \def \showISBNxiii  #1{\unskip}     \fi
\ifx \showISSN     \undefined \def \showISSN      #1{\unskip}     \fi
\ifx \showLCCN     \undefined \def \showLCCN      #1{\unskip}     \fi
\ifx \shownote     \undefined \def \shownote      #1{#1}          \fi
\ifx \showarticletitle \undefined \def \showarticletitle #1{#1}   \fi
\ifx \showURL      \undefined \def \showURL       {\relax}        \fi
\providecommand\bibfield[2]{#2}
\providecommand\bibinfo[2]{#2}
\providecommand\natexlab[1]{#1}
\providecommand\showeprint[2][]{arXiv:#2}

\bibitem[\protect\citeauthoryear{Backstrom, Sun, and Marlow}{Backstrom
  et~al\mbox{.}}{2010}]%
        {backstrom2010find}
\bibfield{author}{\bibinfo{person}{Lars Backstrom}, \bibinfo{person}{Eric Sun},
  {and} \bibinfo{person}{Cameron Marlow}.} \bibinfo{year}{2010}\natexlab{}.
\newblock \showarticletitle{{Find Me if You Can: Improving Geographical
  Prediction with Social and Spatial Proximity}}. In \bibinfo{booktitle}{{\em
  WWW 2010}}. \bibinfo{publisher}{ACM}, \bibinfo{address}{Raleigh, North
  Carolina, USA}, \bibinfo{pages}{61--70}.
\newblock
\showISBNx{978-1-60558-799-8}
\showDOI{%
\url{https://doi.org/10.1145/1772690.1772698}}


\bibitem[\protect\citeauthoryear{Bennett, Radlinski, White, and Yilmaz}{Bennett
  et~al\mbox{.}}{2011}]%
        {bennett2011inferring}
\bibfield{author}{\bibinfo{person}{Paul~N. Bennett}, \bibinfo{person}{Filip
  Radlinski}, \bibinfo{person}{Ryen~W. White}, {and} \bibinfo{person}{Emine
  Yilmaz}.} \bibinfo{year}{2011}\natexlab{}.
\newblock \showarticletitle{{Inferring and Using Location Metadata to
  Personalize Web Search}}. In \bibinfo{booktitle}{{\em SIGIR 2011}}.
  \bibinfo{publisher}{ACM}, \bibinfo{address}{Beijing, China},
  \bibinfo{pages}{135--144}.
\newblock
\showISBNx{978-1-4503-0757-4}
\showDOI{%
\url{https://doi.org/10.1145/2009916.2009938}}


\bibitem[\protect\citeauthoryear{Braden}{Braden}{1989}]%
        {rfc1123}
\bibfield{author}{\bibinfo{person}{R. Braden}.}
  \bibinfo{year}{1989}\natexlab{}.
\newblock \bibinfo{booktitle}{{\em {Requirements for Internet Hosts --
  Application and Support}}}.
\newblock \bibinfo{type}{{RFC}} 1123. \bibinfo{institution}{{RFC Editor}}.
\newblock
\showURL{%
\url{https://tools.ietf.org/html/rfc1123}}


\bibitem[\protect\citeauthoryear{Chabarek and Barford}{Chabarek and
  Barford}{2013}]%
        {chabarek2013s}
\bibfield{author}{\bibinfo{person}{Joseph Chabarek} {and} \bibinfo{person}{Paul
  Barford}.} \bibinfo{year}{2013}\natexlab{}.
\newblock \showarticletitle{What's in a name?: decoding router interface
  names}. In \bibinfo{booktitle}{{\em Proceedings of the 5th ACM workshop on
  HotPlanet}}. ACM, \bibinfo{pages}{3--8}.
\newblock


\bibitem[\protect\citeauthoryear{Chandrasekaran, Bai, Schoenfield, Berger,
  Caruso, Economou, Gilliss, Maggs, Moses, Duff, et~al\mbox{.}}{Chandrasekaran
  et~al\mbox{.}}{2015}]%
        {chandrasekaran2015alidade}
\bibfield{author}{\bibinfo{person}{Balakrishnan Chandrasekaran},
  \bibinfo{person}{Mingru Bai}, \bibinfo{person}{Michael Schoenfield},
  \bibinfo{person}{Arthur Berger}, \bibinfo{person}{Nicole Caruso},
  \bibinfo{person}{George Economou}, \bibinfo{person}{Stephen Gilliss},
  \bibinfo{person}{Bruce Maggs}, \bibinfo{person}{Kyle Moses},
  \bibinfo{person}{David Duff}, {et~al\mbox{.}}}
  \bibinfo{year}{2015}\natexlab{}.
\newblock \bibinfo{booktitle}{{\em Alidade: {IP} geolocation without active
  probing}}.
\newblock \bibinfo{type}{{T}echnical {R}eport}.
  \bibinfo{institution}{Department of Computer Science, Duke University}.
\newblock


\bibitem[\protect\citeauthoryear{Ciavarrini, Greco, and Vecchio}{Ciavarrini
  et~al\mbox{.}}{2018}]%
        {ciavarrini2018geolocation}
\bibfield{author}{\bibinfo{person}{Gloria Ciavarrini}, \bibinfo{person}{Maria~S
  Greco}, {and} \bibinfo{person}{Alessio Vecchio}.}
  \bibinfo{year}{2018}\natexlab{}.
\newblock \showarticletitle{Geolocation of Internet hosts: Accuracy limits
  through Cram{\'e}r--Rao lower bound}.
\newblock \bibinfo{journal}{{\em Computer Networks\/}}  \bibinfo{volume}{135}
  (\bibinfo{year}{2018}), \bibinfo{pages}{70--80}.
\newblock


\bibitem[\protect\citeauthoryear{Dan, Parikh, and Davison}{Dan
  et~al\mbox{.}}{2016}]%
        {dan2016improving}
\bibfield{author}{\bibinfo{person}{Ovidiu Dan}, \bibinfo{person}{Vaibhav
  Parikh}, {and} \bibinfo{person}{Brian~D Davison}.}
  \bibinfo{year}{2016}\natexlab{}.
\newblock \showarticletitle{Improving IP geolocation using query logs}. In
  \bibinfo{booktitle}{{\em Proceedings of the Ninth ACM International
  Conference on Web Search and Data Mining}}. ACM, \bibinfo{pages}{347--356}.
\newblock


\bibitem[\protect\citeauthoryear{Eidnes, de~Groot, and Vixie}{Eidnes
  et~al\mbox{.}}{1998}]%
        {rfc2317}
\bibfield{author}{\bibinfo{person}{H. Eidnes}, \bibinfo{person}{G. de Groot},
  {and} \bibinfo{person}{P. Vixie}.} \bibinfo{year}{1998}\natexlab{}.
\newblock \bibinfo{booktitle}{{\em {Classless IN-ADDR.ARPA delegation}}}.
\newblock \bibinfo{type}{{RFC}} 2317. \bibinfo{institution}{{RFC Editor}}.
\newblock
\showURL{%
\url{https://tools.ietf.org/html/rfc2317}}


\bibitem[\protect\citeauthoryear{Endo and Sadok}{Endo and Sadok}{2010}]%
        {endo2010whois}
\bibfield{author}{\bibinfo{person}{P.T. Endo} {and} \bibinfo{person}{D.
  Sadok}.} \bibinfo{year}{2010}\natexlab{}.
\newblock \showarticletitle{{Whois Based Geolocation: A Strategy to Geolocate
  Internet Hosts}}. In \bibinfo{booktitle}{{\em AINA 2010}}.
  \bibinfo{pages}{408--413}.
\newblock
\showISSN{1550-445X}
\showDOI{%
\url{https://doi.org/10.1109/AINA.2010.39}}


\bibitem[\protect\citeauthoryear{Eriksson, Barford, Sommers, and
  Nowak}{Eriksson et~al\mbox{.}}{2010}]%
        {eriksson2010learning}
\bibfield{author}{\bibinfo{person}{Brian Eriksson}, \bibinfo{person}{Paul
  Barford}, \bibinfo{person}{Joel Sommers}, {and} \bibinfo{person}{Robert
  Nowak}.} \bibinfo{year}{2010}\natexlab{}.
\newblock \showarticletitle{A learning-based approach for IP geolocation}. In
  \bibinfo{booktitle}{{\em International Conference on Passive and Active
  Network Measurement}}. Springer, \bibinfo{pages}{171--180}.
\newblock


\bibitem[\protect\citeauthoryear{for Applied Internet Data~Analysis}{for
  Applied Internet Data~Analysis}{2018}]%
        {ddec2018}
\bibfield{author}{\bibinfo{person}{Center for Applied Internet Data~Analysis}.}
  \bibinfo{year}{2018}\natexlab{}.
\newblock \bibinfo{title}{{DDec - DNS Decoded - CAIDA's public DNS Decoding
  database}}.
\newblock   (\bibinfo{year}{2018}).
\newblock
\showURL{%
\url{http://ddec.caida.org/help.pl}}
\newblock
\shownote{Accessed: 2018-07-31.}


\bibitem[\protect\citeauthoryear{for Europe}{for Europe}{2018}]%
        {unlocode2018}
\bibfield{author}{\bibinfo{person}{United Nations Economic~Commission for
  Europe}.} \bibinfo{year}{2018}\natexlab{}.
\newblock \bibinfo{title}{{UN/LOCODE: United Nations Code for Trade and
  Transport Locations}}.
\newblock   (\bibinfo{year}{2018}).
\newblock
\showURL{%
\url{https://www.unece.org/cefact/locode/welcome.html}}
\newblock
\shownote{Accessed: 2018-06-27.}


\bibitem[\protect\citeauthoryear{Foundation}{Foundation}{2018}]%
        {publicsuffix2018}
\bibfield{author}{\bibinfo{person}{Mozilla Foundation}.}
  \bibinfo{year}{2018}\natexlab{}.
\newblock \bibinfo{title}{Public Suffix List}.
\newblock   (\bibinfo{year}{2018}).
\newblock
\showURL{%
\url{https://publicsuffix.org/list/}}
\newblock
\shownote{Accessed: 2018-06-28.}


\bibitem[\protect\citeauthoryear{Freedman, Vutukuru, Feamster, and
  Balakrishnan}{Freedman et~al\mbox{.}}{2005}]%
        {freedman2005geographic}
\bibfield{author}{\bibinfo{person}{Michael~J Freedman},
  \bibinfo{person}{Mythili Vutukuru}, \bibinfo{person}{Nick Feamster}, {and}
  \bibinfo{person}{Hari Balakrishnan}.} \bibinfo{year}{2005}\natexlab{}.
\newblock \showarticletitle{Geographic locality of IP prefixes}. In
  \bibinfo{booktitle}{{\em Proceedings of the 5th ACM SIGCOMM conference on
  Internet Measurement}}. USENIX Association, \bibinfo{pages}{13--13}.
\newblock


\bibitem[\protect\citeauthoryear{Gharaibeh, Shah, Huffaker, Zhang, Ensafi, and
  Papadopoulos}{Gharaibeh et~al\mbox{.}}{2017}]%
        {gharaibeh2017look}
\bibfield{author}{\bibinfo{person}{Manaf Gharaibeh}, \bibinfo{person}{Anant
  Shah}, \bibinfo{person}{Bradley Huffaker}, \bibinfo{person}{Han Zhang},
  \bibinfo{person}{Roya Ensafi}, {and} \bibinfo{person}{Christos
  Papadopoulos}.} \bibinfo{year}{2017}\natexlab{}.
\newblock \showarticletitle{A look at router geolocation in public and
  commercial databases}. In \bibinfo{booktitle}{{\em Proceedings of the 2017
  Internet Measurement Conference}}. ACM, \bibinfo{pages}{463--469}.
\newblock


\bibitem[\protect\citeauthoryear{Gueye, Ziviani, Crovella, and Fdida}{Gueye
  et~al\mbox{.}}{2006}]%
        {gueye2006constraint}
\bibfield{author}{\bibinfo{person}{Bamba Gueye}, \bibinfo{person}{Artur
  Ziviani}, \bibinfo{person}{Mark Crovella}, {and} \bibinfo{person}{Serge
  Fdida}.} \bibinfo{year}{2006}\natexlab{}.
\newblock \showarticletitle{{Constraint-Based Geolocation of Internet Hosts}}.
\newblock \bibinfo{journal}{{\em IEEE/ACM Transactions on Networking\/}}
  \bibinfo{volume}{14}, \bibinfo{number}{6} (\bibinfo{date}{Dec}
  \bibinfo{year}{2006}), \bibinfo{pages}{1219--1232}.
\newblock
\showISSN{1063-6692}
\showDOI{%
\url{https://doi.org/10.1109/TNET.2006.886332}}


\bibitem[\protect\citeauthoryear{Guo, Liu, Shen, Wang, Yu, and Zhang}{Guo
  et~al\mbox{.}}{2009}]%
        {guo2009mining}
\bibfield{author}{\bibinfo{person}{Chuanxiong Guo}, \bibinfo{person}{Yunxin
  Liu}, \bibinfo{person}{Wenchao Shen}, \bibinfo{person}{H.J. Wang},
  \bibinfo{person}{Qing Yu}, {and} \bibinfo{person}{Yongguang Zhang}.}
  \bibinfo{year}{2009}\natexlab{}.
\newblock \showarticletitle{{Mining the Web and the Internet for Accurate IP
  Address Geolocations}}. In \bibinfo{booktitle}{{\em INFOCOM 2009}}.
  \bibinfo{pages}{2841--2845}.
\newblock
\showISSN{0743-166X}
\showDOI{%
\url{https://doi.org/10.1109/INFCOM.2009.5062243}}


\bibitem[\protect\citeauthoryear{Hannak, Sapiezynski, Molavi~Kakhki,
  Krishnamurthy, Lazer, Mislove, and Wilson}{Hannak et~al\mbox{.}}{2013}]%
        {hannak2013measuring}
\bibfield{author}{\bibinfo{person}{Aniko Hannak}, \bibinfo{person}{Piotr
  Sapiezynski}, \bibinfo{person}{Arash Molavi~Kakhki},
  \bibinfo{person}{Balachander Krishnamurthy}, \bibinfo{person}{David Lazer},
  \bibinfo{person}{Alan Mislove}, {and} \bibinfo{person}{Christo Wilson}.}
  \bibinfo{year}{2013}\natexlab{}.
\newblock \showarticletitle{Measuring personalization of web search}. In
  \bibinfo{booktitle}{{\em Proceedings of the 22nd international conference on
  World Wide Web}}. ACM, \bibinfo{pages}{527--538}.
\newblock


\bibitem[\protect\citeauthoryear{Huang, Maltz, Li, and Greenberg}{Huang
  et~al\mbox{.}}{2011}]%
        {huang2011public}
\bibfield{author}{\bibinfo{person}{Cheng Huang}, \bibinfo{person}{D.A. Maltz},
  \bibinfo{person}{Jin Li}, {and} \bibinfo{person}{Albert Greenberg}.}
  \bibinfo{year}{2011}\natexlab{}.
\newblock \showarticletitle{{Public DNS system and Global Traffic Management}}.
  In \bibinfo{booktitle}{{\em INFOCOM 2011}}. \bibinfo{pages}{2615--2623}.
\newblock
\showISSN{0743-166X}
\showDOI{%
\url{https://doi.org/10.1109/INFCOM.2011.5935088}}


\bibitem[\protect\citeauthoryear{Huffaker, Fomenkov, and claffy}{Huffaker
  et~al\mbox{.}}{2014}]%
        {drop2014}
\bibfield{author}{\bibinfo{person}{B. Huffaker}, \bibinfo{person}{M. Fomenkov},
  {and} \bibinfo{person}{k. claffy}.} \bibinfo{year}{2014}\natexlab{}.
\newblock \showarticletitle{{DRoP:DNS-based Router Positioning}}.
\newblock \bibinfo{journal}{{\em ACM SIGCOMM Computer Communication Review
  (CCR)\/}} \bibinfo{volume}{44}, \bibinfo{number}{3} (\bibinfo{date}{Jul}
  \bibinfo{year}{2014}), \bibinfo{pages}{6--13}.
\newblock


\bibitem[\protect\citeauthoryear{{Internet Assigned Numbers
  Authority}}{{Internet Assigned Numbers Authority}}{2017}]%
        {ianaspprp2017}
\bibfield{author}{\bibinfo{person}{{Internet Assigned Numbers Authority}}.}
  \bibinfo{year}{2017}\natexlab{}.
\newblock \bibinfo{booktitle}{{\em {IANA IPv4 Special-Purpose Address
  Registry}}}.
\newblock \bibinfo{type}{{T}echnical {R}eport}. \bibinfo{institution}{{IANA}}.
\newblock
\showURL{%
\url{https://www.iana.org/assignments/iana-ipv4-special-registry/iana-ipv4-special-registry.xhtml}}
\newblock
\shownote{Accessed: 2018-08-10.}


\bibitem[\protect\citeauthoryear{IP2Location.com}{IP2Location.com}{2018}]%
        {ip2location2018}
\bibfield{author}{\bibinfo{person}{IP2Location.com}.}
  \bibinfo{year}{2018}\natexlab{}.
\newblock \bibinfo{title}{{Geolocate IP Address Location using IP2Location}}.
\newblock   (\bibinfo{year}{2018}).
\newblock
\showURL{%
\url{https://www.ip2location.com/}}
\newblock
\shownote{Accessed: 2018-08-13.}


\bibitem[\protect\citeauthoryear{Jayant and Katz-Bassett}{Jayant and
  Katz-Bassett}{2004}]%
        {jayant2004toward}
\bibfield{author}{\bibinfo{person}{Rika Jayant} {and} \bibinfo{person}{Ethan
  Katz-Bassett}.} \bibinfo{year}{2004}\natexlab{}.
\newblock \bibinfo{booktitle}{{\em Toward Better Geolocation: Improving
  Internet Distance Estimates Using Route Traces}}.
\newblock \bibinfo{type}{{T}echnical {R}eport}. \bibinfo{institution}{The
  Pennsylvania State University}.
\newblock


\bibitem[\protect\citeauthoryear{Katz-Bassett, John, Krishnamurthy, Wetherall,
  Anderson, and Chawathe}{Katz-Bassett et~al\mbox{.}}{2006}]%
        {katz2006towards}
\bibfield{author}{\bibinfo{person}{Ethan Katz-Bassett}, \bibinfo{person}{John~P
  John}, \bibinfo{person}{Arvind Krishnamurthy}, \bibinfo{person}{David
  Wetherall}, \bibinfo{person}{Thomas Anderson}, {and} \bibinfo{person}{Yatin
  Chawathe}.} \bibinfo{year}{2006}\natexlab{}.
\newblock \showarticletitle{Towards IP geolocation using delay and topology
  measurements}. In \bibinfo{booktitle}{{\em Proceedings of the 6th ACM SIGCOMM
  conference on Internet measurement}}. ACM, \bibinfo{pages}{71--84}.
\newblock


\bibitem[\protect\citeauthoryear{Kliman-Silver, Hannak, Lazer, Wilson, and
  Mislove}{Kliman-Silver et~al\mbox{.}}{2015}]%
        {kliman2015location}
\bibfield{author}{\bibinfo{person}{Chloe Kliman-Silver}, \bibinfo{person}{Aniko
  Hannak}, \bibinfo{person}{David Lazer}, \bibinfo{person}{Christo Wilson},
  {and} \bibinfo{person}{Alan Mislove}.} \bibinfo{year}{2015}\natexlab{}.
\newblock \showarticletitle{Location, location, location: The impact of
  geolocation on web search personalization}. In \bibinfo{booktitle}{{\em
  Proceedings of the 2015 Internet Measurement Conference}}. ACM,
  \bibinfo{pages}{121--127}.
\newblock


\bibitem[\protect\citeauthoryear{K{\"o}lmel and Alexakis}{K{\"o}lmel and
  Alexakis}{2002}]%
        {kolmel2002location}
\bibfield{author}{\bibinfo{person}{Bernhard K{\"o}lmel} {and}
  \bibinfo{person}{Spiros Alexakis}.} \bibinfo{year}{2002}\natexlab{}.
\newblock \showarticletitle{Location Based Advertising}. In
  \bibinfo{booktitle}{{\em First International Conference on Mobile Business}}.
  \bibinfo{address}{Athens, Greece}.
\newblock


\bibitem[\protect\citeauthoryear{Laki, M{\'a}tray, H{\'a}ga, Csabai, and
  Vattay}{Laki et~al\mbox{.}}{2010}]%
        {laki2010model}
\bibfield{author}{\bibinfo{person}{S{\'a}ndor Laki}, \bibinfo{person}{P{\'e}ter
  M{\'a}tray}, \bibinfo{person}{P{\'e}ter H{\'a}ga},
  \bibinfo{person}{Istv{\'a}n Csabai}, {and} \bibinfo{person}{G{\'a}bor
  Vattay}.} \bibinfo{year}{2010}\natexlab{}.
\newblock \showarticletitle{A model based approach for improving router
  geolocation}.
\newblock \bibinfo{journal}{{\em Computer Networks\/}} \bibinfo{volume}{54},
  \bibinfo{number}{9} (\bibinfo{year}{2010}), \bibinfo{pages}{1490--1501}.
\newblock


\bibitem[\protect\citeauthoryear{Li, Chen, Guo, Liu, Zhang, Zhang, and
  Zhang}{Li et~al\mbox{.}}{2012}]%
        {li2012ip}
\bibfield{author}{\bibinfo{person}{Dan Li}, \bibinfo{person}{Jiong Chen},
  \bibinfo{person}{Chuanxiong Guo}, \bibinfo{person}{Yunxin Liu},
  \bibinfo{person}{Jinyu Zhang}, \bibinfo{person}{Zhili Zhang}, {and}
  \bibinfo{person}{Yongguang Zhang}.} \bibinfo{year}{2012}\natexlab{}.
\newblock \showarticletitle{IP-Geolocation mapping for moderately-connected
  Internet regions}.
\newblock \bibinfo{journal}{{\em IEEE Transactions on Parallel and Distributed
  Systems\/}} (\bibinfo{year}{2012}).
\newblock


\bibitem[\protect\citeauthoryear{MacVittie}{MacVittie}{2012}]%
        {macvittie2010geolocation}
\bibfield{author}{\bibinfo{person}{Lori MacVittie}.}
  \bibinfo{year}{2012}\natexlab{}.
\newblock \bibinfo{title}{{Geolocation and Application Delivery}}.
\newblock
  \bibinfo{howpublished}{\url{https://www.f5.com/pdf/white-papers/geolocation-wp.pdf}}.
    (\bibinfo{year}{2012}).
\newblock
\newblock
\shownote{Accessed: 2018-08-02.}


\bibitem[\protect\citeauthoryear{MaxMind}{MaxMind}{2018}]%
        {maxmind2018}
\bibfield{author}{\bibinfo{person}{Inc. MaxMind}.}
  \bibinfo{year}{2018}\natexlab{}.
\newblock \bibinfo{title}{{Detect Online Fraud and Locate Online Visitors}}.
\newblock   (\bibinfo{year}{2018}).
\newblock
\showURL{%
\url{https://www.maxmind.com/en/home}}
\newblock
\shownote{Accessed: 2018-08-13.}


\bibitem[\protect\citeauthoryear{Mockapetris}{Mockapetris}{1987}]%
        {rfc1034}
\bibfield{author}{\bibinfo{person}{P. Mockapetris}.}
  \bibinfo{year}{1987}\natexlab{}.
\newblock \bibinfo{booktitle}{{\em {DOMAIN NAMES - CONCEPTS AND FACILITIES}}}.
\newblock \bibinfo{type}{{RFC}} 1034. \bibinfo{institution}{{RFC Editor}}.
\newblock
\showURL{%
\url{https://tools.ietf.org/html/rfc1034}}


\bibitem[\protect\citeauthoryear{Muir and Oorschot}{Muir and Oorschot}{2009}]%
        {muir2009internet}
\bibfield{author}{\bibinfo{person}{James~A. Muir} {and} \bibinfo{person}{Paul
  C.~Van Oorschot}.} \bibinfo{year}{2009}\natexlab{}.
\newblock \showarticletitle{Internet geolocation: Evasion and counterevasion}.
\newblock \bibinfo{journal}{{\em ACM Computing Surveys (CSUR)\/}}
  \bibinfo{volume}{42}, \bibinfo{number}{1} (\bibinfo{year}{2009}),
  \bibinfo{pages}{4}.
\newblock


\bibitem[\protect\citeauthoryear{{Neustar, Inc.}}{{Neustar, Inc.}}{2018}]%
        {neustar2018}
\bibfield{author}{\bibinfo{person}{{Neustar, Inc.}}}
  \bibinfo{year}{2018}\natexlab{}.
\newblock \bibinfo{title}{{IP Intelligence}}.
\newblock   (\bibinfo{year}{2018}).
\newblock
\showURL{%
\url{https://www.security.neustar/digital-performance/ip-intelligence}}
\newblock
\shownote{Accessed: 2018-08-13.}


\bibitem[\protect\citeauthoryear{NUR and TOZAL}{NUR and TOZAL}{2018}]%
        {2018nurgeography}
\bibfield{author}{\bibinfo{person}{ABDULLAH~YASIN NUR} {and}
  \bibinfo{person}{MEHMET~ENGIN TOZAL}.} \bibinfo{year}{2018}\natexlab{}.
\newblock \showarticletitle{{Geography and Routing in the Internet}}.
\newblock \bibinfo{journal}{{\em ACM Transactions on Spatial Algorithms and
  Systems (TSAS)\/}} (\bibinfo{year}{2018}).
\newblock


\bibitem[\protect\citeauthoryear{P.~Faltstrom}{P.~Faltstrom}{2003}]%
        {rfc3490}
\bibfield{author}{\bibinfo{person}{A.~Costello P.~Faltstrom, P.~Hoffman}.}
  \bibinfo{year}{2003}\natexlab{}.
\newblock \bibinfo{booktitle}{{\em {Internationalizing Domain Names in
  Applications (IDNA)}}}.
\newblock \bibinfo{type}{{RFC}} 3490. \bibinfo{institution}{{RFC Editor}}.
\newblock
\showURL{%
\url{https://tools.ietf.org/html/rfc3490}}


\bibitem[\protect\citeauthoryear{Padmanabhan and Subramanian}{Padmanabhan and
  Subramanian}{2001}]%
        {padmanabhan2001investigation}
\bibfield{author}{\bibinfo{person}{Venkata~N. Padmanabhan} {and}
  \bibinfo{person}{Lakshminarayanan Subramanian}.}
  \bibinfo{year}{2001}\natexlab{}.
\newblock \showarticletitle{{An Investigation of Geographic Mapping Techniques
  for Internet Hosts}}. In \bibinfo{booktitle}{{\em SIGCOMM 2001}}.
  \bibinfo{publisher}{ACM}, \bibinfo{address}{San Diego, California, USA},
  \bibinfo{pages}{173--185}.
\newblock
\showISBNx{1-58113-411-8}
\showDOI{%
\url{https://doi.org/10.1145/383059.383073}}


\bibitem[\protect\citeauthoryear{Poese, Uhlig, Kaafar, Donnet, and Gueye}{Poese
  et~al\mbox{.}}{2011}]%
        {poese2011ip}
\bibfield{author}{\bibinfo{person}{Ingmar Poese}, \bibinfo{person}{Steve
  Uhlig}, \bibinfo{person}{Mohamed~Ali Kaafar}, \bibinfo{person}{Benoit
  Donnet}, {and} \bibinfo{person}{Bamba Gueye}.}
  \bibinfo{year}{2011}\natexlab{}.
\newblock \showarticletitle{IP geolocation databases: Unreliable?}
\newblock \bibinfo{journal}{{\em ACM SIGCOMM Computer Communication Review\/}}
  \bibinfo{volume}{41}, \bibinfo{number}{2} (\bibinfo{year}{2011}),
  \bibinfo{pages}{53--56}.
\newblock


\bibitem[\protect\citeauthoryear{Rainie and Duggan}{Rainie and Duggan}{2016}]%
        {rainie2016privacy}
\bibfield{author}{\bibinfo{person}{Lee Rainie} {and} \bibinfo{person}{Maeve
  Duggan}.} \bibinfo{year}{2016}\natexlab{}.
\newblock \showarticletitle{Privacy and information sharing}.
\newblock \bibinfo{journal}{{\em Pew Research Center\/}}  \bibinfo{volume}{16}
  (\bibinfo{year}{2016}).
\newblock


\bibitem[\protect\citeauthoryear{Rapid7Labs}{Rapid7Labs}{2017}]%
        {rapid7rdns2013}
\bibfield{author}{\bibinfo{person}{Rapid7Labs}.}
  \bibinfo{year}{2017}\natexlab{}.
\newblock \bibinfo{title}{Reverse DNS (RDNS) - 2013-2017}.
\newblock
  \bibinfo{howpublished}{\url{https://opendata.rapid7.com/sonar.rdns/}}.
  (\bibinfo{year}{2017}).
\newblock
\newblock
\shownote{Accessed: 2018-06-23.}


\bibitem[\protect\citeauthoryear{Rapid7Labs}{Rapid7Labs}{2018}]%
        {rapid7rdns2018}
\bibfield{author}{\bibinfo{person}{Rapid7Labs}.}
  \bibinfo{year}{2018}\natexlab{}.
\newblock \bibinfo{booktitle}{{\em Reverse DNS (RDNS) v2 - 2017 onward}}.
\newblock \bibinfo{type}{{T}echnical {R}eport}.
\newblock
\showURL{%
\url{https://opendata.rapid7.com/sonar.rdns_v2/}}
\newblock
\shownote{Accessed: 2018-06-23.}


\bibitem[\protect\citeauthoryear{Scheitle, Gasser, Sattler, and Carle}{Scheitle
  et~al\mbox{.}}{2017}]%
        {scheitle2017hloc}
\bibfield{author}{\bibinfo{person}{Quirin Scheitle}, \bibinfo{person}{Oliver
  Gasser}, \bibinfo{person}{Patrick Sattler}, {and} \bibinfo{person}{Georg
  Carle}.} \bibinfo{year}{2017}\natexlab{}.
\newblock \showarticletitle{HLOC: Hints-Based Geolocation Leveraging Multiple
  Measurement Frameworks}.
\newblock \bibinfo{journal}{{\em arXiv preprint arXiv:1706.09331\/}}
  (\bibinfo{year}{2017}).
\newblock


\bibitem[\protect\citeauthoryear{Shavitt and Zilberman}{Shavitt and
  Zilberman}{2011}]%
        {shavitt2011geolocation}
\bibfield{author}{\bibinfo{person}{Yuval Shavitt} {and} \bibinfo{person}{Noa
  Zilberman}.} \bibinfo{year}{2011}\natexlab{}.
\newblock \showarticletitle{A geolocation databases study}.
\newblock \bibinfo{journal}{{\em IEEE Journal on Selected Areas in
  Communications\/}} \bibinfo{volume}{29}, \bibinfo{number}{10}
  (\bibinfo{year}{2011}), \bibinfo{pages}{2044--2056}.
\newblock


\bibitem[\protect\citeauthoryear{Shue, Paul, and Taylor}{Shue
  et~al\mbox{.}}{2013}]%
        {Shue2013}
\bibfield{author}{\bibinfo{person}{Craig~A. Shue}, \bibinfo{person}{Nathanael
  Paul}, {and} \bibinfo{person}{Curtis~R. Taylor}.}
  \bibinfo{year}{2013}\natexlab{}.
\newblock \showarticletitle{{From an IP Address to a Street Address: Using
  Wireless Signals to Locate a Target}}. In \bibinfo{booktitle}{{\em WOOT
  2013}}. \bibinfo{publisher}{USENIX}, \bibinfo{address}{Washington, D.C.}
\newblock
\showURL{%
\url{https://www.usenix.org/conference/woot13/workshop-program/presentation/Shue}}


\bibitem[\protect\citeauthoryear{Spring, Mahajan, and Wetherall}{Spring
  et~al\mbox{.}}{2002}]%
        {undns2002}
\bibfield{author}{\bibinfo{person}{Neil Spring}, \bibinfo{person}{Ratul
  Mahajan}, {and} \bibinfo{person}{David Wetherall}.}
  \bibinfo{year}{2002}\natexlab{}.
\newblock \showarticletitle{Measuring ISP topologies with Rocketfuel}.
\newblock \bibinfo{journal}{{\em ACM SIGCOMM Computer Communication Review\/}}
  \bibinfo{volume}{32}, \bibinfo{number}{4} (\bibinfo{year}{2002}),
  \bibinfo{pages}{133--145}.
\newblock


\bibitem[\protect\citeauthoryear{Svantesson}{Svantesson}{2007}]%
        {svantesson2007commerce}
\bibfield{author}{\bibinfo{person}{Dan Jerker~B Svantesson}.}
  \bibinfo{year}{2007}\natexlab{}.
\newblock \showarticletitle{{E-Commerce Tax: How The Taxman Brought Geography
  To The `Borderless' Internet}}.
\newblock \bibinfo{journal}{{\em Revenue Law Journal\/}} \bibinfo{volume}{17},
  \bibinfo{number}{1} (\bibinfo{year}{2007}), \bibinfo{pages}{11}.
\newblock


\bibitem[\protect\citeauthoryear{Timmins}{Timmins}{2018}]%
        {telcodata2018}
\bibfield{author}{\bibinfo{person}{Paul Timmins}.}
  \bibinfo{year}{2018}\natexlab{}.
\newblock \bibinfo{title}{TelcoData Telecommunications Database}.
\newblock   (\bibinfo{year}{2018}).
\newblock
\showURL{%
\url{https://www.telcodata.us/}}
\newblock
\shownote{Accessed: 2018-06-27.}


\bibitem[\protect\citeauthoryear{Wang, Burgener, Flores, Kuzmanovic, and
  Huang}{Wang et~al\mbox{.}}{2011}]%
        {wang2011towards}
\bibfield{author}{\bibinfo{person}{Yong Wang}, \bibinfo{person}{Daniel
  Burgener}, \bibinfo{person}{Marcel Flores}, \bibinfo{person}{Aleksandar
  Kuzmanovic}, {and} \bibinfo{person}{Cheng Huang}.}
  \bibinfo{year}{2011}\natexlab{}.
\newblock \showarticletitle{{Towards Street-level Client-independent IP
  Geolocation}}. In \bibinfo{booktitle}{{\em NSDI 2011}}.
  \bibinfo{publisher}{USENIX}, \bibinfo{address}{Berkeley, CA, USA},
  \bibinfo{pages}{365--379}.
\newblock
\showURL{%
\url{http://dl.acm.org/citation.cfm?id=1972457.1972494}}


\bibitem[\protect\citeauthoryear{Wick}{Wick}{2018}]%
        {geonames2018}
\bibfield{author}{\bibinfo{person}{Marc Wick}.}
  \bibinfo{year}{2018}\natexlab{}.
\newblock \bibinfo{title}{GeoNames}.
\newblock   (\bibinfo{year}{2018}).
\newblock
\showURL{%
\url{http://download.geonames.org/export/dump/}}
\newblock
\shownote{Accessed: 2018-06-27.}


\bibitem[\protect\citeauthoryear{Wong, Stoyanov, and Sirer}{Wong
  et~al\mbox{.}}{2007}]%
        {wong2007octant}
\bibfield{author}{\bibinfo{person}{Bernard Wong}, \bibinfo{person}{Ivan
  Stoyanov}, {and} \bibinfo{person}{Emin~G\"{u}n Sirer}.}
  \bibinfo{year}{2007}\natexlab{}.
\newblock \showarticletitle{{Octant: A Comprehensive Framework for the
  Geolocalization of Internet Hosts}}. In \bibinfo{booktitle}{{\em NSDI 2007}}.
  \bibinfo{publisher}{USENIX Association}, \bibinfo{address}{Berkeley, CA,
  USA}, \bibinfo{pages}{23--23}.
\newblock
\showURL{%
\url{http://dl.acm.org/citation.cfm?id=1973430.1973453}}


\bibitem[\protect\citeauthoryear{Youn, Mark, and Richards}{Youn
  et~al\mbox{.}}{2009}]%
        {youn2009statistical}
\bibfield{author}{\bibinfo{person}{I. Youn}, \bibinfo{person}{B.L. Mark}, {and}
  \bibinfo{person}{D. Richards}.} \bibinfo{year}{2009}\natexlab{}.
\newblock \showarticletitle{{Statistical Geolocation of Internet Hosts}}. In
  \bibinfo{booktitle}{{\em ICCCN 2009}}. \bibinfo{pages}{1--6}.
\newblock
\showISSN{1095-2055}
\showDOI{%
\url{https://doi.org/10.1109/ICCCN.2009.5235373}}


\end{thebibliography}
%
%
\end{document}